\documentclass[aps,prd,nofootinbib,superscriptaddress, showkeys,english,twocolumn,superscriptaddress]{revtex4-2}

\usepackage{babel}
\usepackage{graphicx}
\usepackage{color}
\usepackage{amsmath}
\usepackage{amssymb}
\usepackage{multirow}
\usepackage{color}
\usepackage{comment}
\usepackage{appendix}
\usepackage[normalem]{ulem}
\usepackage{hyperref}
\usepackage{float} % en el preámbulo
\usepackage{cancel}

\begin{document}
\title{Two-Step Bose–Einstein Condensation of an ideal Magnetized  Charged Bosonic gas under neutron star-like conditions}

	\author{Amanda Castillo Ay\'on}
	\email{amandac@icimaf.cu}
	\affiliation{Instituto de Cibern\'{e}tica, Matem\'{a}tica y F\'{\i}sica (ICIMAF), \\
		Calle E esq a 15 Vedado 10400 La Habana Cuba}
 \author{Gabriel Gil Perez}
  \affiliation {Instituto de Cibern\'{e}tica, Matem\'{a}tica y F\'{\i}sica (ICIMAF), \\
		Calle E esq a 15 Vedado 10400 La Habana Cuba}
	\author{A. P\'erez Mart\'inez\footnote{Instituto de Cibern\'etica Matem\'atica y F\'isica (ICIMAF), Cuba}}
	\email{aurora@icimaf.cu}
	\author{H. P\'erez Rojas}
	\email{hugo@icimaf.cu}
	\affiliation {Instituto de Cibern\'{e}tica, Matem\'{a}tica y F\'{\i}sica (ICIMAF), \\
		Calle E esq a 15 Vedado 10400 La Habana Cuba}
  \author{Gabriella Piccinelli Bocchi}
	\email{itzamna@unam.mx}
	\affiliation{Centro Tecnológico, FES Aragón-UNAM, México\\
    Av. Universidad Nacional S/N, Bosques de Aragón, 57171 Cdad.
Nezahualcóyotl, Mexico.
}
          \author{Adriel Rodr\'iguez Concepci\'on}
	\email{adrielerdz@icimaf.cu}
	\affiliation{Instituto de Cibern\'{e}tica, Matem\'{a}tica y F\'{\i}sica (ICIMAF), \\
		Calle E esq a 15 Vedado 10400 La Habana Cuba}
  \author{Angel Sanchez}
	\email{ansac@ciencias.unam.mx}
	\affiliation{Facultad de Ciencias, Universidad Nacional Aut\' onoma de M\' exico, Apartado Postal 50-542, Ciudad de M\'exico 04510, M\'exico.}
	\date{\today}
\date{\today}

%%%%%%%%%%%%%%%%%%%%%%%%%%%%%%%%%%%%%%%%%%%%%%%%%%%%%%%%%%%%%%%%%%%%%%%%%%%%%%%%%%%%%%%%%%%%%%%%%%%%%%%%%%%%%%%%%%%%%%%%%%%%%

\begin{abstract}
We study the thermodynamic properties of a non-interacting, relativistic gas of charged scalar bosons in a uniform magnetic field, including both statistical and vacuum contributions at arbitrary field strengths. Focusing on the low-temperature regime and separating the Lowest Landau Level (LLL) from excited states provides a clearer view of the magnetic field’s impact on thermodynamic quantities.
We revisit Bose–Einstein condensation (BEC), specific heat, magnetization, and the  equation of state (EoS). A central result is the diffuse character of BEC induced by the magnetic field, reflected in the specific heat, which exhibits two plateaus: the first appears when the system becomes effectively one-dimensional through magnetic confinement, while the second—associated with the true one-particle ground state—is suppressed by the field. Consequently, no critical condensation temperature arises.
For magnetization, the LLL contribution shifts from diamagnetic to paramagnetic As the field strengthen with the inclusion of the excited states, the statistical magnetization remains negative. In contrast, the vacuum contribution dominates at strong fields, driving paramagnetism. We also show that antiparticles enhance specific heat, magnetization, and total pressure.
These effects are illustrated for a pion gas under neutron star conditions and compared with previous results for a neutral vector boson system.

\end{abstract}

%%%%%%%%%%%%%%%%%%%%%%%%%%%%%%%%%%%%%%%%%%%%%%%%%%%%%%%%%%%%%%%%%%%%%%%%%%%%%%%%%%%%%%%%%%%%%%%%%%%%%%%%%%%%%%%%%%%%%%%%%%%%%

\maketitle

\section{Introduction}

Alongside brilliant stars, the Universe is also home to stellar remnants—what we might call “stellar corpses”—whose nature depends on the characteristics of the stars they originated from. When a star reaches the end of its life, it often ends in a dramatic collapse followed by a bright supernova explosion. Stars with intermediate masses ($8 M_{\odot} \leq M \leq 20 M_{\odot}$) typically become neutron stars (NS). The name has historical origins: the first explanation for the stability of these incredibly dense objects suggested that they were supported by the pressure of a degenerate gas of massive fermions. At the time, scientists proposed that these fermions were neutrons, which had just been discovered \cite{Chadwick1932}.

However, it is already known that the observational data cannot be fully explained by this composition alone. Instead, a wide range of possible constituents have been proposed, ranging from neutrons to quarks in different phases and subjected to different interactions. In addition, phase transitions to a condensate or a superfluid of bosons are also considered in some models.

 The high density is one of the most important features of NS with masses on the order of that of our Sun enclosed in a region of roughly $10$ km, the density of these objects can reach that of atomic nuclei ($\sim 10^{14}$ g/cm$^3$).
No less important is their precise and rapid rotation, which led to their discovery \cite{Hewish1968}. If they possess a magnetic field misaligned with their rotation axis, NS can emit significant amounts of energy as electromagnetic waves due to the precession of the magnetic field. 
Most of this radiation falls within the range of radio waves, X-rays, and gamma rays, with only a very small fraction in the visible range of the electromagnetic spectrum.

Observations of NS have revealed several remarkable features. According to observational data, these stars possess ultrahigh magnetic fields at their surfaces, around $10^{12}-10^{13}$~G in radio pulsars and reaching $10^{15}$~G in magnetars \cite{Kouveliotou1992}. Nevertheless, there are still many uncertainties about these magnetic fields, including their origin and the intensities allowed in the stars' inner regions.

Matter subjected to these extreme conditions of density, temperature, and magnetic field has not been recreated in terrestrial laboratories yet, leaving the validation of the true composition of NS to astronomical observations. Research in this field evolves in two directions: the construction of theoretical models to compare with observations and the imposition of astrophysical constraints on these models, shedding light on high-energy physics. In the "sea" of theoretical models, the possibility of condensates of pions or superfluid states of protons, neutrons or pions is not excluded, and we are interested in exploring them in presence of magnetic fields.

Although our primary motivation for this study arises from astrophysical applications, the results may also provide insight into Bose-Einstein condensates (BECs) of neutral atoms and molecules \cite{Delgado2012Twostep}. In particular, recent experiments with dipolar bosonic molecules—such as NaCs—featuring long-range dipole-dipole interactions have demonstrated tunability from weak to strong coupling regimes. This level of control has opened the door to the exploration of exotic quantum phases, including supersolids, dipolar crystals, and Mott insulators with fractional fillings \cite{Bigagli_2023, Bigagli_2024}.

The magnetized ideal charged scalar boson gas (MCSBG) has previously been studied in both relativistic and non-relativistic regimes, and remains an active topic in the literature. Among the most relevant works are \cite{Delgado2012Twostep, ROJAS1996148, ROJAS1997,perez1999boseeinsteincondensationconstantmagnetic, Daicic1996, Khalilov1997, Khalilov:1999xd, Standen1999} and references therein, which focus on understanding whether a critical temperature for Bose–Einstein condensation (BEC) exists in the presence of a magnetic field. 
Summarizing the main findings, in the relativistic case, condensation and magnetization have been analyzed separately in the weak–field (WF) and strong–field (SF) regimes \cite{ROJAS1996148, Khalilov1997, Khalilov:1999xd, ROJAS1997, perez1999boseeinsteincondensationconstantmagnetic}. In the WF limit, it was argued that the gas undergoes a conventional BEC transition, with the critical temperature depending on the magnetic field. In the SF regime, however, all particles are confined to the lowest Landau level (LLL), making the system effectively one-dimensional. In this case, a critical temperature cannot be defined, but there exists an interval of temperatures at which the bosons concentrate around the ground state, indicating the occurrence of a diffuse phase transition \cite{ROJAS1996148, ROJAS1997}. In \cite{Daicic1996}, a precise analysis of magnetization, in both relativistic and non-relativistic regimes, was performed and the existence of conventional condensation in arbitrary dimensions was examined. 
In \cite{Standen1999}, exact expressions for the specific heat and magnetization, in the non-relativistic case were obtained, and the possibility of conventional condensation by introducing a definition of a critical temperature based on the maximum of the specific heat was discussed. On the other hand, in Ref.~\ \cite{Delgado2012Twostep}, the two-step condensation through the behavior of the specific heat and particle clustering was investigated for a non-relativistic system. 
For the ideal MCSBG, in Ref.~\cite{Ayala:2012dk}, by introducing self-interactions in the charged boson gas, a well-defined critical temperature and a conventional BEC was obtained. The authors further argue that the magnetic field catalyzes the condensate. 

Related to Bose Einstein condensate but for the case of a magnetized neutral vector boson gas (MNVBG) are the works  \cite{Quintero2017PRC, QuinteroAngulo2021, PhysRevC.108.015806, Angulo:2017hif, Concepcion:2024duj} where the thermodynamic behavior was investigated at arbitrary temperatures, and the corresponding equation of state (EoS) was derived with the goal of modeling Bose-Einstein Condensate stars as alternatives to conventional neutron stars. The analysis revealed two distinct regimes depending on the temperature. At low temperatures, magnetic field effects dominate, giving rise to spontaneous magnetization, pressure anisotropy, and the possibility of transverse magnetic collapse. At high temperatures, however, the system is governed by pair production, which restores pressure isotropy and significantly enhances both magnetization and total pressure. These results highlight the importance of relativistic and thermal effects in extreme astrophysical environments, where even neutral bosonic systems can exhibit nontrivial magnetic responses.

In this paper, within the Scalar-QED framework, we extend this line of research by analyzing the MCSBG, aiming to explore its thermodynamic properties and EoS in the context of our ongoing research on BEC stars. Our initial attempts in this direction are presented in \cite{Angulo:2023lil}, where some thermodynamic properties are discussed, and more recently in  \cite{Hernandez:2025bvn}, where neutron stars with charged magnetized pions in their core are studied. We start from the relativistic charged boson propagator in the one loop approximation at finite temperature and density and in presence of a magnetic field to derive the thermodynamic potential. From it, we compute the thermodynamic properties of the system under an arbitrary magnetic field by separating the contribution from particles in the Lowest Landau Level (LLL) from those in higher Landau levels, following the approach inspired by \cite{Ayala:2012dk}, where BEC with self-interactions was analyzed. Separating the LLL contribution allows us to properly deal with the strong infrared divergences in the particle density. 

The novelty of our work, compared to the studies mentioned above, lies in three key aspects. First, in the approach itself,  which, through the calculation of a relativistic and exact analytical expression for the thermodynamical potential, allows for the investigation of the low-temperature regime at arbitrary values of the magnetic field, enabling a systematic exploration of the interplay between temperature and magnetic field in the system. Second, in the new results and physical insights that emerge as a direct consequence of this approach, ranging from the analysis of the two-step Bose–Einstein condensation to the derivation of the thermodynamics properties of the system, as the anisotropic equations of state  and the system transition from diamagnetic to paramagnetic states. Third, in the explicit inclusion of vacuum contributions and antiparticles, which are often neglected in previous studies. 
In particular, we confirm that condensation effectively occurs as a two-step process, consistent with the findings of \cite{Delgado2012Twostep} in the non-relativistic case. Furthermore, we uncover the dual role of the magnetic field in this two-step condensation: on the one hand, it catalyzes the first step by reducing the dimensionality of the system, while on the other hand, it inhibits the second step, which corresponds to reaching the ground state.  
With this comprehensive study, we establish the groundwork for the subsequent investigation of Bose–Einstein condensate stars.

The article is organized as follows. In Section \ref{thermo}, we compute the thermodynamic potential of the relativistic MCSBG. In section \ref{LT section} we derive the low-temperature limit and \ref{sec::BEC} is devoted to the Bose-Einstein condensation. The magnetization of the gas is obtained and discussed in Section \ref{sec:magn}, and in Section \ref{sec:EoS} we study the EoS. In Section \ref{conclusions}, we summarize the results and present our conclusions. In addition, we present appendixes that complement our discussions. 

Although we have astrophysical motivations and illustrate our results with pions of mass $m_{\pi}\sim 140$ MeV, we use dimensionless units, so that our conclusions can be applied, for any charged scalar boson gas in other context like heavy ions colliders  or condensate matter physics. Throughout this paper we  use natural units, where $k_B = c = \hbar = \dfrac{1}{4\pi\epsilon_0} = 1$.

%%%%%%%%%%%%%%%%%%%%%%%%%%%%%%%%%%%%%%%%%%%%%%%%%%%%%%%%%%%%%%%%%%%%%%%%%%%%%%%%%%%%%%%%%%%%%%%%%%%%%%%%%%%%%%%%%%%%%%%%%%%%%%%

\section{Thermodynamic potential}
\label{thermo}

We consider the thermodynamic potential of a charged boson gas in the presence of a uniform and constant magnetic field in the $z$-direction in the one-loop approximation. In the framework of the imaginary-time formalism, the expression for the thermodynamic potential reads:

\begin{align}
\Omega(T,\mu,B)&=\frac{eB T}{4\pi^2} \sum_{n,l}\int dp_3 \ln{{ D_l^{-1}(p^*)}},
\label{OmegaT}
\end{align}

\noindent where $p^*=(i\omega_n-\mu,0,\sqrt{(2l+1)eB},p_3)$ and the summation over $n$ runs over the Matsubara frequencies,  $T$ is the absolute temperature, $\mu$ and $e$ are the bosons chemical potential and charge, respectively, and $l$ stands for the Landau Levels quantization of the transverse momentum to the magnetic field. The integrand in Eq. \ref{OmegaT} is the inverse boson Green function in the presence of a magnetic field (see appendix \eqref{functionaleffective}, for details) which reads:
\begin{equation}
D_l^{-1}(p^*)= ((i\omega_n-\mu)^2+p_3^2+(2l+1)eB +m^2).
\end{equation}
 
 Once the sum over the Matsubara frequencies is performed, the thermodynamic potential can be split into two contributions: the well-known \textit{vacuum} contribution $\Omega_{V}(B)$ that depends solely on the magnetic field intensity and requires renormalization; and the so called \textit{statistical} contribution $\Omega_{\text{st}}(T, \mu, B)$, that depends on the temperature, chemical potential and magnetic field intensity. This last term can in turn be split into the LLL and the contribution of the excited states yielding
 
\begin{widetext}
\begin{align}\label{Eq1}
\Omega(T,\mu,B) &=\Omega_{V}(B)+\Omega_{st}(T,\mu,B) \nonumber \\
&= \frac{eB}{4\pi^2}\sum_{l=0}^{\infty} \int_{-\infty}^{\infty} E_l(p_3) dp_3  +\frac{eB}{4\pi^2 \beta}\left (\int_{-\infty}^{\infty} dp_3 \ln [(1-e^{-\beta(E_0(p_3)-\mu)})(1-e^{-\beta(E_0(p_3)+\mu)})]\right.\nonumber\\
&\hspace{2em} +\left.\sum_{l=1}^{\infty}\int_{-\infty}^{\infty} dp_3 \ln [(1-e^{-\beta(E_l(p_3)-\mu)})(1-e^{-\beta(E_l(p_3)+\mu)})]\right ),
\end{align}
\end{widetext}

\noindent
where $\beta=1/T$ is the inverse of the absolute temperature. The particle energy spectrum is $E_l(p_3) = \sqrt{p^2_3+m_B^2+2eBl}$, with $m_B=\sqrt{m^2+eB}$ the boson effective mass, that includes a magnetic field contribution. In particular, for the LLL, $E_0(p_3)= \sqrt{p^2_3+m_B^2}$.

In the last equation, we have separated the LLL term from the rest of the thermal contribution, as is commonly done for fermions \cite{Ferrer2015AMM}. For charged fermions, in the strong magnetic field regime the confinement of particles in the LLL leads to a dimensional reduction --refereed to as \textit{magnetic catalysis} \cite{Gusynin:1995nb}-- of the system and gives rise to various interesting phenomena. One such phenomenon occurs in vacuum, where the magnetic field catalyzes the formation of a fermion condensate \cite{Gusynin:1995nb}. As we will see later, a similar dimensional reduction takes place for charged bosons in a magnetic field, also known as magnetic catalysis \cite{Ayala:2012dk}, with important implications for the collective behavior that drives Bose-Einstein condensation. The phenomena under discussion, occurring in both fermionic and bosonic systems, ultimately stem from the breaking of the $SO(3)$ symmetry due to the presence of a magnetic field \cite{Chaichian:1999gd}.

By using a standard procedure~\cite{Schwinger1951}, the sum over Landau levels and the integral over $p_3$ in the vacuum contribution can be straightforwardly performed, obtaining:
\begin{equation}
 \Omega_{\text{V}}(B)
= -\frac{eB}{16\pi^2} \int_{0}^{\infty} \frac{ds}{s^2} \dfrac{ e^{-m^2 s}}{\sinh(eBs)}
\label{vacBpropertime}.
\end{equation}

Since the expression above contains the vacuum contribution and the vacuum polarization, both need to be isolated and regularized \cite{Schwinger1951}, as shown in appendix~\ref{potreg}. The finite contribution to the vacuum thermodynamic potential reads Eq. \eqref{Omegarenor} where $\zeta$ is the Hurwitz zeta function and the dimensionless magnetic field variable $b\equiv B/B_c$, with $B_c=m^2/e$, was introduced. In the last line in Eq.(\ref{Omegarenor}), the $\epsilon$-technique was used~\cite{W.Dittrich46}.
 
\begin{figure*}[ht]
 \begin{align}
     \Omega^R_{V}(B) &= -\frac{1}{16\pi^2} \int_{0}^{\infty} \frac{ds}{s^3} \left(\dfrac{eBs}{\sinh(eBs)} -1 + \frac{(eBs)^2}{6}\right) e^{-m^2 s}\\
     &= -\frac{(eB)^{2}}{16\pi^2}
     \left[
     -\frac{\gamma_E-\ln b}{6}+\frac{1}{4b^2}(-3+2\gamma_E-\ln b^2)  +4(-1+\gamma_E+\ln2)\zeta\left(-1,\frac{b+1}{2b}\right) -4\zeta^{(1,0)}\left(-1,\frac{b+1}{2b}\right)\right]  ,   
     \label{Omegarenor}
 \end{align}
\end{figure*}
 
In thermodynamic studies of particle properties, the regularized vacuum is typically neglected. However, this term contributes to the vacuum energy $\Omega^R_V(B)$ as zero-point energy, which plays a crucial role in the magnetic properties of the gas, and consequently in the equation of state.

%%---------------------------------------------------------------------------------------------------------------------------
\subsection{Any-temperature thermodynamic potential}

Let us start by considering the statistical part of the thermodynamic potential for the MCSBG in Eq(3) for arbitrary values of the magnetic field,  at finite temperature and density, given by

\begin{align}
    \Omega_{st}(&T,\mu,B) = \dfrac{e B}{4\pi^2 \beta} \sum_{l=0}^\infty \int_{-\infty}^\infty dp_3 \times \nonumber \\
    &\ln \left[ (1 - e^{\beta(\mu - E_l(p_3)} )  (1 - e^{-\beta(\mu + E_l(p_3)} )    \right].
\label{Omega_st_inicial_proper_time}
\end{align}

\noindent 
In the above equation, by using a Taylor series for the logarithm \footnote{$\ln(1-x) = -\sum_{k = 1}^{\infty}\frac{x^k}{k}$ for $|x| < 1$.  This is valid for the MCSBG since $e^{\beta(\mu - E_l(p_3))} < 1$ for every energy level and finite temperature, as proved in \cite{ROJAS1996148} and by us in Sect. \ref{sec::BEC}.},  together with the identity
\begin{align}
e^{-\beta n \sqrt{p_3^2 + 2eB(l+1/2)+m^2}} &= \frac{\beta n}{2} \int_{0}^{\infty}\frac{ds}{\sqrt{\pi} s^{3/2}} \times \nonumber \\
&e^{-\frac{\beta^2 n^2}{4s}-s (p_3^2 + 2eB(l+1/2)+m^2)},
\label{identitypropertime}
\end{align}
the integral over $p_3$ and the sum over Landau levels can be straightforwardly performed, yielding \cite{Ayala:2012dk}
\begin{align}
\Omega_{st}(&T,\mu,B)=-\frac{eB}{4\pi^2}\int_{0}^{\infty} \frac{ds}{s^2}e^{-m_B^2s} \times \nonumber \\
&\left( 1+ \frac{1}{e^{2eBs}-1}\right) \sum^{\infty}_{n=1} e^{-\frac{n^2 \beta^2}{4s}} \cosh(\mu \beta n),
\end{align}

\noindent
where the parameter $s$ can be related to Schwinger's proper time.

The first term inside the parenthesis is the contribution of the LLL and the second one corresponds to the excited Landau levels contributions, denoted as $\Omega_{LLL}$ and $\Omega_{l\ne 0}$, respectively, namely
\textbf{\begin{subequations}
\begin{align}
\Omega_{st}(&T,\mu,B)=\Omega_{LLL} (T,\mu,B)+\Omega_{l\ne 0}(T,\mu,B),\label{OmegaTsep}\\
\Omega_{LLL}(&T,\mu,B)=-\frac{eB}{4\pi^2}\int_{0}^{\infty} \frac{ds}{s^2}e^{-m_B^2s} \times \nonumber \\
&\left( \sum^{\infty}_{n=1} e^{-\frac{n^2 \beta^2}{4s}} \cosh(\mu \beta n)\right),\\
\Omega_{l\ne 0}(&T,\mu,B)=-\frac{eB}{4\pi^2}\int_{0}^{\infty} \frac{ds}{s^2}e^{-m_B^2s}\times \nonumber \\ 
&\frac{1}{e^{2eBs}-1} \left( \sum^{\infty}_{n=1} e^{-\frac{n^2 \beta^2}{4s}} \cosh(\mu \beta n)\right).
\label{OmegasepaL}
\end{align}
\end{subequations}
}

These expressions are valid for arbitrary values of the magnetic field and temperature. 

The contribution of the LLL to the dimensionless thermodynamic potential, $\hat{\Omega}
\equiv \Omega/m^4$,  can be rewritten as
\begin{equation} \hat{\Omega}_{LLL}= - \dfrac{bt m_b}{\pi^2} \sum_{n=1}^\infty \dfrac{1}{n} \cosh \left(\dfrac{nx}{t}\right)
 K_1\left(\dfrac{n m_b}{t}\right),
 \label{omegaLLL}
\end{equation}
where the dimensionless ground-state energy $m_b\equiv m_B/m=\sqrt{1+b}$,  temperature $t\equiv T/m$ and  chemical potential $x\equiv \mu/m$,  have been introduced, together with the integral representation of the modified Bessel functions of the second kind, given by \cite{Arfken:2005}
\[
     K_\nu\left(z\right)=\dfrac{1}{2}\left(\dfrac{z}{2}\right)^\nu \int_0^\infty e^{-t - \frac{z^2}{4t}}\dfrac{dt}{t^{\nu+1}},
\]
for $|\text{arg}(z)| < \pi/4$. It is worth to note that this result was also obtained in Ref.\cite{Angulo:2023lil}. The key difference lies in the summation over Landau levels, where we explicitly separate the LLL in Eq. (\ref{OmegaTsep}) from the sum over the excited Landau levels.

The contribution of the excited Landau Levels to the thermodynamic potential,  Eq. (\ref{OmegasepaL}),  can  be written in terms of a modified Bessel function only in the case  $b\ll t$, where it takes the form
\begin{equation}
\hat{\Omega}_{l\ne 0}\approx m^{2}_bt^2\sum_{n=1}^\infty \frac{1}{n^2} K_2\left(\dfrac{nm_b}{t}\right)\cosh \left(\dfrac{nx}{t}\right).\label{WFHT}
\end{equation}
This result was also obtained  in Ref. \cite{Khalilov:1999xd} for the weak field limit, which manifests the equivalence between the weak-field and high-temperature limits.
 
As we will discuss later, the LLL is the leading term in the thermodynamic quantities at low temperatures (as compared to the magnetic field), and its infrared divergences are responsible for the diffuse condensation of MCSBGs, as pointed out in \cite{ROJAS1996148, perez1999boseeinsteincondensationconstantmagnetic, Delgado2012Twostep}. Therefore, isolating this term from the rest is convenient for an analytical and numerical treatment of those divergences, contributing to a better understanding of the thermodynamic properties of the MCSBG.

%%%%%%%%%%%%%%%%%%%%%%%%%%%%%%%%%%%%%%%%%%%%%%%%%%%%%%%%%%%%%%%%%%%%%%%%%%%%%%%%%%%%%%%%%%%%%%%%%%%%%%%%%%%%%%%%%%%%%%%%%%%%%
 
\section{Low temperature regime: thermodynamic properties}\label{LT section}

In this section, we derive several thermodynamic quantities of interest in the low-temperature regime, characterized by temperatures well below the ground-state energy ($t\ll m_b$), and we work with dimensionless quantities~\footnote{We use the notation $\hat{X}\equiv X/m^a$, with $a$ a number such that $\hat{X}$ becomes a dimensionless thermodynamic quantity.}.

To explore the thermodynamic properties of the MCSBG in the low temperature regime, we shall resort
to the Laplace method which allows us to approximate the integral over $s$ in Eqs. (\ref{OmegaTsep})-(\ref{OmegasepaL}), in the limit $m_b/t\rightarrow\infty$, as follows~\cite{Arfken:2005}:
\begin{subequations}
\begin{align}
\int_{0}^{\infty} ds \ g(s)e^{\frac{m_b}{t}f(s)} \sim \frac{\sqrt{2\pi t}\, g(s_0)}{\mid m_b  f^{''}(s_0)\mid^{1/2}}e^{\frac{m_b}{t}f(s_0)}.
\end{align}
\end{subequations}
where $s_0$ is a maximum point of the function $f(s)$, which in our case is  $f(s)=-s- \frac{n^2}{4s}$ and  $s_0=n/2$. With this approximation one finds that the dimensionless thermodynamic potential, in the low temperature regime $(LT)$, behaves as
\begin{subequations}
\begin{align}
\hat{\Omega}^{LT}_{st} &=  \hat{\Omega}^{+,LT}_{st} + \hat{\Omega}^{-,LT}_{st}\\
\hat{\Omega}^{\pm,LT}_{st} &= \hat{\Omega}^{\pm,LT}_{LLL}(t,x,b) + \Omega_{l\ne 0}^{\pm,LT}(t,x,b),\\
\hat{\Omega}^{\pm,LT}_{LLL}(t,x,b) &= - b  m_b^{1/2}
    \left(\dfrac{t}{2\pi}\right)^{3/2}Li_{3/2}(z_\pm), \label{Omegalowt} \\
\hat{\Omega}^{\pm,LT}_{l\ne 0}(t,x,b) &=
- b  m_b^{1/2}
    \left(\dfrac{t}{2\pi}\right)^{3/2}\displaystyle \sum_{n=1}^\infty\dfrac{z_\pm^n}{n^{3/2}}\dfrac{1}{e^{n\gamma}-1}.  
\end{align}
\end{subequations}

\noindent
The notation \(^{\pm}\) indicates that antiparticles are included in the calculations: the \((+)\) refers to the particle contribution, and the \((-)\) to the antiparticle contribution. We define \(z_\pm \equiv e^{(\pm x - m_b)/ t}\), where $z_+$ the gas fugacity; and \(\gamma \equiv b/(t m_b)\), as the scaled magnetic field. The thermodynamic potential $\hat{\Omega}^{LT}$ is valid for arbitrary values of the magnetic field\footnote{From now on $t\ll m_b$ and the "LT" label will be dropped in the thermodynamic magnitudes.}, and from now on, we refer to it as the exact result.

The LLL term coincides with $\hat{\Omega}_{st}$ in the strong-field approximation \cite{Khalilov1997}, whereas for the $l\ne 0$ terms a weak field approximation, $b \ll t \ll m_{b}$, can be considered, obtaining
\begin{equation}
\hat{\Omega}_{l\ne 0}^\pm(t,x,b)  \approx -  t  \left( \dfrac{m_b t}{2 \pi} \right)^{3/2}  Li_{5/2}(z_\pm).
\end{equation}

%%%----------------------------------------------------------------------------------------------------------------------------

\subsection{Particle number and energy density.}

With the thermodynamic potential, we can compute all the thermodynamic properties. In particular, the gas density, defined as 
\begin{equation}
\hat{N}= - \left( \dfrac{\partial\hat{\Omega}}{\partial x} \right)_{t,b},
\end{equation}
is the difference between the particle density $\hat{N}^+$ and the antiparticle density $\hat{N}^-$,   
\begin{equation}
\hat{N}(t,x,b) = \hat{N}^+(t,x,b)  - \hat{N}^-(t,x,b),
\label{density}
\end{equation}
where
\begin{subequations}
\begin{align}
\hat{N}^\pm(t,x,b)& =  \hat{N}^\pm_{LLL}(t,x,b)  + \hat{N}^\pm_{l\ne 0}(t,x,b), \\
\hat{N}^\pm_{LLL}(t,x,b)
&=\frac{b (m_b t)^{1/2}}{(2\pi)^{3/2}}Li_{1/2}(z_\pm),\\
\hat{N}^{\pm}_{l\ne 0}(t,x,b)&=
\frac{b (m_b t)^{1/2}}{(2\pi)^{3/2}}\displaystyle \sum_{n=1}^\infty\frac{z_\pm^n}{n^{1/2}}\frac{1}{e^{n\gamma}-1}.
\end{align} \label{pnumber}
\end{subequations}

The direct proportionality between $\hat{N}_{LLL}^+$ and $Li_{1/2}(z_+)$ leads to significant consequences. Indeed, since $\displaystyle\lim_{z \rightarrow 1^-} Li_{1/2}(z) = +\infty$ the fugacity must satisfy $z_+ < 1$ at any finite temperature to allow $\hat{N}$ to remain bounded. As we discuss in the next section, this condition determines the type of Bose-Einstein condensation exhibited by MCSBGs  \cite{ROJAS1996148,perez1999boseeinsteincondensationconstantmagnetic,Ayala:2012dk}. Moreover, by fixing the density $\hat{N}$ to a specific value $\rho$, one obtains the dependence of the dimensionless chemical potential $x$ on the density, temperature and magnetic field through the implicit equation $\hat{N}(t,x(t,\rho,b),b) = \rho$. 

As before, in the regime $m_b \gg t \gg b$, the $l\ne0$ contribution can be approximated as 
 \begin{equation}
 \hat{N}^\pm_{l\ne 0}(t,x,b) \approx \left(\frac{m_b t}{2\pi} \right)^{3/2}Li_{{3/2}}(z_{\pm}) .
 \label{Naprox}
 \end{equation}

The dimensionless statistical energy density, defined as $\hat{E}_{st}= \hat{\Omega}_{st}+ t\hat{S}+x\hat{N}$, with the dimensionless entropy density  $\hat{S} = -\left({\partial\hat{\Omega}}/{\partial t}\right)_{x,b}$, receives contributions from both particles and antiparticles,
\begin{equation}
\hat{E}_{st} = \hat{E}_{st}^+ + \hat{E}_{st}^-,
\label{energy}
\end{equation}
with
\begin{subequations}
\begin{align}
\hat{E}_{st}^\pm(&t,x,b) = \hat{E}_{LLL}^\pm(t,x,b) +\hat{E}_{l\ne 0}^\pm(t,x,b),\\
\hat{E}_{LLL}^{\pm}(&t,x,b)= \dfrac{b (m_b t)^{1/2} }{2(2\pi)^{3/2}} \Bigg[ 2m_bLi_{1/2}(z_\pm)  \nonumber \\   &+t Li_{{3/2}}(z_\pm)\Bigg]\\
\hat{E}_{l\ne 0}^\pm(&t,x,b)=
\dfrac{b}{(2\pi)^{3/2}} \left( \dfrac{t}{m_b} \right)^{1/2} \times \nonumber \\
&\displaystyle \sum_{n=1}^\infty\dfrac{ z_\pm^n}{n^{{3/2}} (e^{n\gamma} - 1)^2}  \Bigg[ -n m_b^2  \nonumber \\ 
&+(1+2b)ne^{n\gamma} +\frac{1}{2}m_bt (e^{n\gamma} -1)\Bigg]. 
\label{energy_density}
\end{align}
\end{subequations}

\noindent
In the regime $b \ll t \ll m_b$, the contributions of the excited Landau levels can be approximated by
\begin{align}
    \hat{E}_{l\ne 0}^\pm(&t,x,b) \approx \left(\dfrac{m_b t}{2\pi}\right)^{3/2} \left(\right.
        \dfrac{3}{2} t \, Li_{5/2}(z) \nonumber \\ & + m_b \, Li_{3/2}(z)
    \left.\right)
    \label{Eaprox}
\end{align}

\begin{figure}[h!]
    \centering    \includegraphics[width=.9\linewidth]{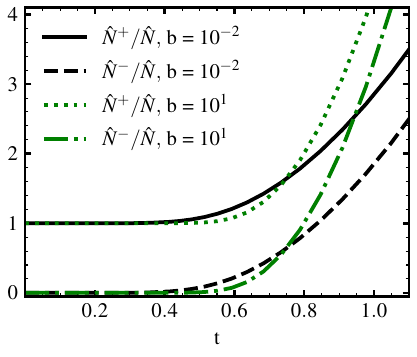}
    \caption{Relative fraction of particles and antiparticles in the system as a function of temperature and for two values of the magnetic field and fixed $\hat{N} = 0.012$.}
    \label{fig:Np_Nm}
\end{figure}

In order to asses the role of antiparticles in the thermodynamic properties of the gas let us analyze the particle and antiparticle densities as functions of temperature at fixed magnetic field values and a fixed density $\hat{N} = 0.012$. \footnote{
    For a pion gas, this density corresponds to $\sim 10^{12}$ g/cm$^3$, which can be found inside neutron stars, where densities can be as high as $\sim 10^{14}$ g/cm$^3$ \cite{Shapiro}. At the same time, this density allows the LT approximation to be used in the temperature range in which the transition to the BEC is observed.
}

In Fig.~\ref{fig:Np_Nm}, we show the system’s behavior for a weak and a strong field intensity. The antiparticle density starts to rapidly increase around $t \sim 0.4-0.6$ and, at temperatures around  $t \sim 0.6-0.7$, the antiparticle fraction reaches nearly half of the density, with $\hat{N}^- \gtrsim \frac{\hat{N}}{2}$. Furthermore, the figure shows that higher magnetic fields increase the antiparticle fraction. As a result, even in the low-temperature regime, antiparticles must be taken into account when analyzing the thermodynamic properties of the boson gas. They could only be neglected in the study of Bose-Einstein condensation, since this phenomenon occurs at even lower temperatures, as $t \to 0$.

%%%%%%%%%%%%%%%%%%%%%%%%%%%%%%%%%%%%%%%%%%%%%%%%%%%%%%%%%%%%%%%%%%%%%%%%%%%%%%%%%%%%%%%%%%%%%%%%%%%%%%%%%%%%%%%%%%%%%%%%%%%%%%

\section{Bose-Einstein condensation}\label{sec::BEC}

Having obtained the relevant thermodynamic quantities of the MCSBG, we now turn to one of the central goals of this work: the analysis of its Bose–Einstein condensation. Following the approach of \cite{ROJAS1996148}, we employ two different criteria to classify the type of BEC that occurs: \textit{strong} and \textit{weak}. 

The strong criterion posits the existence of a critical temperature $t_c$ such that, if the gas is cooled at fixed density, the fugacity reaches $z = 1$ for $t = t_c$ and remains the same at lower temperatures. As a consequence, a macroscopic fraction of bosons accumulates in the ground state of the system at $t <t_c$ and, as $t \to 0$, every boson falls into the ground state. In contrast, the weak criterion applies when $z < 1$ for all $t > 0$, implying that no finite critical temperature can be defined. In this scenario, a macroscopic fraction of the particles accumulates not only in the ground state but also in its immediate vicinity as $t \to 0$.

When the strong criterion holds, the condensation is referred to as \textit{usual} and a phase transition into the Bose-Einstein condensate is observed as the system temperature decreases at fixed particle density \cite{ROJAS1996148}. For a non-magnetized boson gas in three dimensions, the phase transition to the BEC is of second order \cite{ROJAS1996148} and it is manifested as a peak in the plot of the specific heat per particle at $t = t_c$ \cite{Pathria}. In App. \eqref{App Bec b = 0} we include more details about the usual BEC.

As discussed previously, the MCSBG exhibits a divergence in the particle density and, as a consequence, $z < 1$ at any finite temperature, and a critical temperature cannot be defined. The weak criterion thus applies in this case and the phase transition to the BEC is referred to as \textit{diffuse}. In particular, there is no peak in the specific heat \cite{perez1999boseeinsteincondensationconstantmagnetic, Ayala:2012dk, ROJAS1996148, ROJAS1997}.

We examine the signatures and characteristics of the condensation process within the MCSBG using two complementary approaches. First, in Subsection~\ref{BEC cv}, we derive the specific heat at constant volume and analyze its dependence on the temperature, keeping the particle density and magnetic field fixed. Its behavior allows us to contrast the diffuse BEC found in the MCSBG with the usual BEC observed in non-magnetized systems. Then, in Subsection~\ref{BEC populations}, we study the accumulation of particles in the lowest Landau level and in the vicinity of the ground state and how this connects with the specific heat behavior, gaining further insight into the condensation mechanism. In both approaches, we highlight the two-step condensation process and the magnetic field's influence on the crossover temperatures related to each step. Also, because BEC is an ultra-low-temperature phenomenon, we neglect the presence of antiparticles throughout the remainder of this section\footnote{So $z \equiv z_+$.}.

%%---------------------------------------------------------------------------------------------------------------------------

\subsection{Specific heat of the MCSBG} \label{BEC cv}

Let's now analyze the temperature dependence of the specific heat at constant volume (from now on specific heat) and the imprints of the diffuse BEC. The specific heat is defined as:
\begin{align} 
C_V &\equiv \left( \dfrac{\partial \hat{E}}{\partial t} \right)_{\hat{N},b} = \underbrace{\left( \dfrac{\partial \hat{E}}{\partial t} \right)_{x,b}}_{C_V^T} + \left( \dfrac{\partial x}{\partial t} \right)_{\hat{N},b} \underbrace{\left( \dfrac{\partial \hat{E}}{\partial x} \right)_{t,b}}_{C_V^{\mu}}, \label{CVt}
\end{align} 
with
\begin{align}
\left(\dfrac{\partial x}{\partial t} \right)_{\hat{N},b} &= -\left( \dfrac{\partial \hat{N}}{\partial t} \right)_{x,b} \left( \dfrac{\partial \hat{N}}{\partial x} \right)^{-1}_{t,b}, \label{cv_eq} 
\end{align}
where the three magnitudes $C_V^T$, $C_V^\mu$ and $\left( \dfrac{\partial x}{\partial t} \right)_{\hat{N},b}$ depend on $t$, $x$ and $b$. In our calculation of $C_V$ we consider constant $\hat{N}$ and we fix $b$.

\begin{figure*}[ht]
\begin{subequations}
\begin{align}
    C_V^{T}&\equiv C_V^{T, LLL}+C_V^{T,l\neq0}, \\
   C_V^{T,LLL}&= \frac{b (m_b t)^{1/2} }{4 (2\pi) ^{3/2} } \left[ - 2 Li_{1/2}(z) \ln z +3 Li_{{3/2}}(z)\right], \\
   C_V^{T , l\neq0}&= \dfrac{b}{4(2\pi t m_b)^{3/2}} \sum_{n=1}^\infty  \frac{z^n}{n^{3/2}\left(e^{\gamma  n}-1\right)^3}[ 8 b^2 n^2 e^{2 \gamma  n}-2 n m_b \left(e^{\gamma  n}-1\right) \left(m_b^2 t-e^{\gamma  n} (-2 b n x+3 b t+t)\right), 
    \nonumber \\
    &+b\left(e^{\gamma  n}-1\right) \left(e^{\gamma  n} \left(4 n^2-2 n t x+3 t^2\right)+t (2 n x-3 t)\right)+t \left(e^{\gamma  n}-1\right)^2 (3 t-2 n x)], \nonumber\\
    C_V^{\mu} &\equiv C_V^{\mu,LLL}+C_V^{\mu,l\ne 0},\\
    C_V^{\mu,LLL}&= \frac{b  (m_b t)^{1/2}}{2 (2 \pi )^{3/2}} Li_{1/2}(z),\\
    C_V^{\mu,l\ne 0}&= \dfrac{b (m_b t)^{1/2}}{2(2\pi)^{3/2}} \sum_{n = 1 }^\infty \frac{  z^n }{n^{1/2}} \dfrac{  e^{n\gamma} (1 + 2 \gamma n )- 1}{(e^{n\gamma}-1)^2}
         \end{align} 
   \label{Cv mu T}         
\end{subequations}
\end{figure*}

The explicit forms of $C_V^T$ and $C_V^\mu$ are given by Eqs. \eqref{Cv mu T} and, for $\left( {\partial x}/{\partial t} \right)_{\hat{N},b}$, we have
\begin{widetext}
\begin{equation}
    \left( \dfrac{\partial x}{\partial t} \right)_{\hat{N},b}  =  - \frac{ Li_{1/2}(z) - 2Li_{-1/2}(z) \ln z  + \sum\limits_{n=1}^{\infty} \dfrac{z^n}{n^{1/2}} 
\dfrac{2 n \gamma e^{n\gamma} + (e^{n\gamma} - 1)(1 - 2n \ln z)}{(e^{n\gamma} - 1)^2}
}{2 Li_{-1/2}(z) + \sum\limits_{n=1}^{\infty} \dfrac{z^n \sqrt{n}}{e^{n\gamma} - 1}
}.
    \label{x_derivative}
\end{equation}
\end{widetext}

Note that although we are able to separate the LLL contribution in $C_V^T$ and $C_V^\mu$ we cannot do so in $\left( {\partial x}/{\partial t} \right)_{\hat{N},b}$, since this derivative receives contributions from every Landau level in both numerator and denominator. 

The specific heat given by Eq. \eqref{CVt}, once Eqs.~(\ref{Cv mu T}) together with \eqref{x_derivative} are used, provides us with an expression valid for an arbitrary magnetic field intensity, which from now on we refer to as \textit{exact}, to distinguish it from the approximations we also present.

In strong magnetic field regimes, it is meaningful to maintain only the LLL contribution and drop the contributions from the higher levels since they are exponentially suppressed. In such cases, the specific heat, and the specific heat per particle,  d to: 
\begin{align}
C_{V}^{LLL} 
 = \dfrac{b (m_b t)^{1/2}}{4(2\pi)^{3/2}}\left[-\dfrac{Li^2_{1/2}(z)}{Li_{-1/2}(z)} + 3 Li_{3/2}(z)\right ],
    \label{cv_LLL}
\end{align}
and 
\begin{equation}
\frac{ C_{V}^{LLL}}{\hat{N}_{LLL}} = \dfrac{3}{4} \dfrac{Li_{{3/2}}(z)}{Li_{1/2}(z)} - \dfrac{1}{4} \dfrac{Li_{1/2}(z)}{Li_{-1/2}(z)},\label{CvbLLL}
\end{equation}
resembling the behavior of a one-dimensional gas with mass $m \to m_b$ (see Eqs.(2.7)-(2.8) in \cite{Standen1999} for the case $D=1$).

On the other hand, in the regime $m_b \gg t \gg b$, we can use the energy and particle densities, Eqs. \eqref{density} and \eqref{energy}, along with the approximate expressions in Eqs. \eqref{Naprox} and \eqref{Eaprox}, respectively. In this regime, the specific heat behaves like:
\begin{widetext}
\begin{equation}
  \frac{C_V}{\hat{N}} \approx \dfrac{
  (m_b t)^{5/2}[-9 Li^2_{3/2}(z) + 15 Li_{1/2}(z) Li_{5/2}(z)] + 
  b A_1 +
  b^2 A_2
  }{
  4 (2\pi)^{3/2}[ t m_b Li_{1/2}(z)+ b Li_{-1/2}(z) ]
  },
  \label{cv_approx}
\end{equation}
\end{widetext}
\noindent where
\begin{align*}
    A_1(t,z,b) &= (m_b t)^{3/2}[-3 Li_{1/2}(z) Li_{3/2}(z) \\ &+ 15 Li_{-1/2}(z) Li_{5/2}(z)], \\
    A_2(t,z,b) &= (m_b t)^{1/2}[- Li^2_{1/2}(z) + 3 Li_{-1/2}(z) Li_{3/2}(z)] .
\end{align*}

One can expand Eq. \eqref{cv_approx} in a Taylor series around $b = 0$ obtaining, to first-order: 
\begin{align}
\dfrac{C_V}{\hat{N}} &\approx  \left.\dfrac{C_V}{\hat{N}} \right|_{b=0}+ 
\dfrac{b}{t} \Bigg( -\frac{15 Li_{1/2}(z) Li_{5/2}(z)}{8 Li_{3/2}(z)^2} \nonumber \\ 
&+ \frac{9 Li_{3/2}(z) Li_{-1/2}(z)}{8 Li_{1/2}(z)^2}+\frac{3}{4} \Bigg).
    \label{cv taylor expansion}
\end{align}
Expression \eqref{cv taylor expansion} allows to make the limit $b \to 0$ and recover the well known expression for the free ideal non-magnetized bosonic gas at $t > t_c$ (see the appendix \ref{App Bec b = 0})\cite{Pathria}.

\begin{figure*}[ht!]
\centering
\begin{tabular}{cc}
\includegraphics[width=0.4\linewidth]{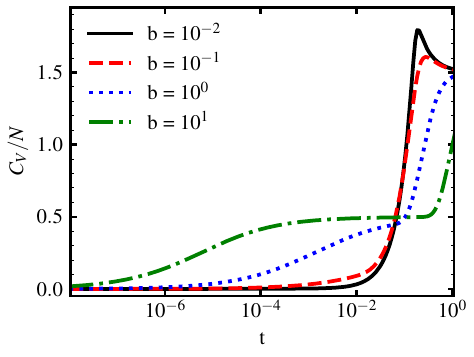} & 
\includegraphics[width=0.4\linewidth]{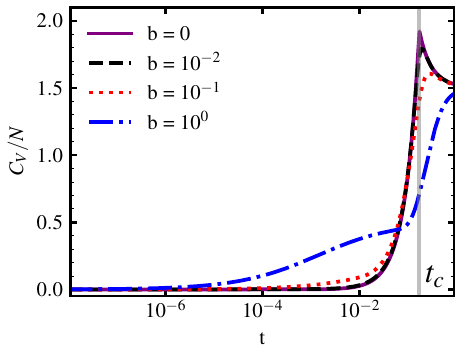}  \\
(a) & (b) \\
\includegraphics[width=0.4\linewidth]{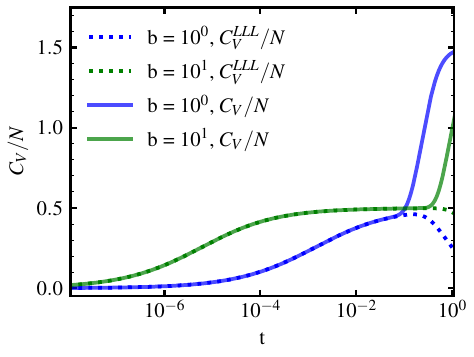} & 
\includegraphics[width=0.4\linewidth]{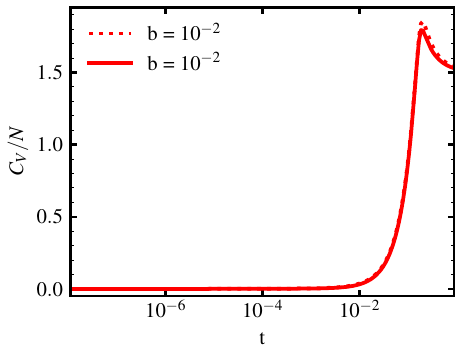}\\
(c) & (d)
\end{tabular}
\caption{
In Fig.(a) the specific heat $C_V/\hat{N}$ is plotted as a function of the temperature for a constant density  $\hat{N} = 0.012$ and four values of the magnetic field.
Fig. (b) includes the $C_V/\hat{N}$ for $B=0$ case (Eq. \eqref{Cv_b0}) to show the  second-order phase transition, indicated by the peak, and to compare it with the smoothed maxima arising for a finite magnetic field strength.
Fig. (c) shows how the specific heat is dominated by the LLL in the presence of a strong magnetic field, computed using Eq. \eqref{cv_LLL}.
Fig.(d) shows the comparison between $C_V/\hat{N}$, computed exactly with Eqs. \eqref{CVt} (solid line), and its low magnetic field approximation given by Eq. \eqref{cv_approx} (dotted line). 
}
\label{fig:cvapp}
\end{figure*}

In Fig.~\ref{fig:cvapp}, we have plotted the specific heat as a function of temperature at different magnetic field intensities and fixed particle density. In particular, in Fig.~\ref{fig:cvapp}a, $C_V/\hat{N}$ is plotted using Eq. (\ref{CVt}). The curves reveal the existence of a local maximum in $C_V/\hat{N}$ for $b = 10^{-2}$ and $b = 10^{-1}$. Also, for $b=10^0$ and $b = 10^1$, the emergence of two plateaus or noticeable shoulders in the behavior of $C_V/\hat{N}$ is observed.

In Fig.~\ref{fig:cvapp}b we have included the zero magnetic field curve, given by Eq. \eqref{Cv_b0}, to show the existence of the  second-order phase transition to the BEC, signaled by the peak at the critical temperature. At $\hat{N} = 0.012$ the critical temperature is approximately $0.17$, which is $\sim 10^{11}$ K for a pion gas.
 As is shown in the plots, this peak softens and disappears when a magnetic field is applied, becoming a local differentiable maximum. For $b = 10^0$ and $b = 10^1$, there are no local maxima in $C_V/\hat{N}$,  being this  always below $3/2$ as shown in Fig.~\ref{fig:cvapp}a. The smoothness of the specific heat curves corroborates the absence of a second-order phase transition as expected from a diffuse phase transition.

In turn, Fig.~\ref{fig:cvapp}c is focused on the plateaus at low temperatures for $b = 10^0$ and $b = 10^1$. We have included the curves of the LLL contribution to the specific heat given by Eq. \eqref{CvbLLL}. The superposition of the curves of $C_V/\hat{N}$ and $C_V^{LLL}/\hat{N}$ in the temperature range where the plateaus and shoulders emerge indicates that the leading contribution to the specific heat in this temperature range comes from the LLL, that is, $C_V^{LLL}$. The 0-plateau corresponds to the accumulation of a macroscopical amount of particles into the ground state and its neighborhood, and $C_V/\hat{N} \to 0$ when $t\to 0$; while the $1/2$-plateaus show the accumulation of particles in the LLL with thermal motion maintained along the magnetic field direction. In this regime, the gas behaves as an effective one-dimensional system with $C_V/\hat{N} \sim 1/2$. As the magnetic field increases, the 0-plateau spans over a narrower temperature range, whereas the width of the $1/2$-plateau increases. As highlighted in \cite{Delgado2012Twostep}, two crossover temperatures appear: one when $C_V/\hat{N}$ leaves the 0-plateaus, and the other when $C_V/\hat{N}$ leaves the $1/2$-plateaus (moving from left to right). 

The existence of these plateaus illustrates the two-step BEC of the MCSBG: first, as the temperature is decreased, \textit{i.e.} moving from right to left in the plots, the gas experiences a dimensional reduction from three dimensions to effectively just one and reaches the 1/2-plateaus (or shoulder).  Since the entrance to this plateau occurs at higher temperatures for a stronger magnetic field, then we could say that the magnetic field stimulates the the first step of the BEC. Conceptually, we may explain this observation in terms of the proportionality of the first Landau excitation energy and the magnetic field strength. As the field increases, the energy gap between the LLL and higher Landau levels widens, making thermal excitation into these excited levels increasingly unlikely. 
Second, as the temperature is further reduced, the diffuse BEC is completely reached at the 0-plateau, although this step is now inhibited by the magnetic field. This inhibition occurs because the magnetic field disfavors the occupation of energy levels near the ground state, as we will prove in detail in section \eqref{BEC populations}.

Finally, Fig.~\ref{fig:cvapp}d compares the specific heat in the limit $t\gg b$, Eq.~\eqref{cv_approx}, with the exact result, Eq.~(\ref{CVt}), for a weak magnetic field. The approximation is generally acceptable, since it reproduces the qualitative behavior of the specific heat, including the limits $C_V/\hat{N} \to 0$ when $t \to 0$ and the classical limit $C_V/\hat{N} \to3/2$ when $z \to 0$. However, for $t\sim b$, it slightly overestimates the maximum value of $C_V/\hat{N}$.

\begin{figure}[h!]
\includegraphics[width=.9\linewidth]{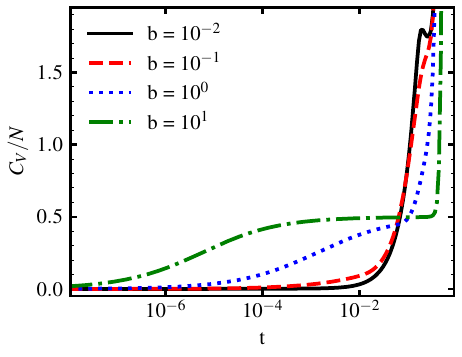}
\caption{The figure shows the specific heat $C_V/\hat{N}$ as a function of the temperature for a density $\hat{N} = 0.012$ and four magnetic field strengths, considering the presence of antiparticles.}
\label{fig:cvanti}
\end{figure}

At last, in order to assess the relevance of antiparticles in thermodynamic properties even at low temperatures, in Fig. (\ref{fig:cvanti}) we have plotted the specific heat, taking into account the presence of antiparticles. The figure highlights how, at temperatures above $\sim 0.1$, all curves rise well above $3/2$, indicating the relativistic behavior of the gas.

%%-----------------------------------------------------------------------------------------------------------------------------

\subsection{Particles population and diffuse BEC}\label{BEC populations}

Let us now delve deeper into the mechanism underlying the diffuse BEC by analyzing the particle density in the vicinity of the system's ground state, given by $|p_3|\le p_0, l = 0$, being  $p_0$ an arbitrarily chosen constant with dimensions of momentum \footnote{In the following plots, we take  the dimensionless momentum $ \hat{p}_0 \equiv p_0/m = 10^{-3}$.}.
We write the $\hat{N}_{LLL}$, Eq.(\ref{pnumber}b), as the sum of two contributions:  
\begin{equation}
\hat{N}_{LLL} =\hat{N}_{gs}+\hat{N}_{k}, \label{pgsk}
\end{equation}
where $\hat{N}_{gs}$ accounts for the particles in such a neighborhood and $\hat{N}_{k}$, for the rest of particles in the LLL, namely those with $l = 0, |p_3| > p_0$.  

Let us calculate $\hat{N}_{gs}$, from the thermodynamic potential in Eq.(\ref{Omega_st_inicial_proper_time}), restricting the integral over $p_3$ to the interval $-p_0 \le p_3 \le p_0$ with $l = 0$.
Using the identity in Eq.~(\ref{identitypropertime}) and performing the integral over $s$, we obtain the contribution to the thermodynamic potential of the particles in such a neighborhood of the ground state as:
\[
\hat{\Omega}_{gs} = - \dfrac{b \, m_b^{1/2} }{\sqrt{2}(\pi \beta)^{{3/2}}} \sum_{n=1}^\infty  \dfrac{\cosh(n x/t)}{n^{{3/2}}} \text{erf}(\alpha \sqrt{n})  e^{-n\beta m_b},
\]
where $\text{erf}(x)$ is the Gaussian error function~\cite{Arfken:2005} and $\alpha = \hat{p}_0/(2m_b t)^{1/2}$. Then, neglecting the antiparticles we get
\begin{equation}
    \hat{N}_{gs} =  \dfrac{b \, (m_b t)^{1/2}}{(2\pi)^{{3/2}}} \sum_{n=1}^\infty  \dfrac{z^n}{n^{1/2}} \text{erf}(\alpha \sqrt{n}).
    \label{particles at gs}
\end{equation}
In the limit \( t \rightarrow 0 \), at fix $b$ (\( \alpha \rightarrow \infty \)), the error function approaches unity, turning the summation in \( \hat{N}_{gs} \) into a polylogarithmic function of order \( 1/2 \), independently of the specific value of \( \hat{p}_0 \). Consequently, \( \hat{N}_{gs} \) becomes equivalent to \( \hat{N}_{LLL} \), as given by Eq.~(\ref{pnumber}b), and $N_k=0$.
Moreover, in this same limit, the total particle density given in Eq.~(\ref{pnumber}a) also reduces to \( N_{LLL} \), implying that \( \hat{N}_{gs} \rightarrow \hat{N} \) as \( t \rightarrow 0 \). This means that, for any arbitrarily small neighborhood of the ground state, there exists a temperature range in which all the particles are contained in that neighborhood.
This behavior is characteristic of the diffuse BEC and was already demonstrated in \cite{ROJAS1996148,ROJAS1997, perez1999boseeinsteincondensationconstantmagnetic} for MCSBGs and one-dimensional gases.

On the other hand, at very low fixed temperature, since 
$\alpha$ is inversely proportional to $m_b^{1/2}$  the increase of the magnetic field  decreases $\hat{N}_{gs}$.
In this sense, the magnetic field inhibits the condensation.
Physically, this means that the population of particles near the true
ground state increases
as the gas is cooled, but decreases as the magnetic field strength
increases.

\begin{figure*}[ht!]
\centering
\begin{tabular}{cc}
\includegraphics[width=0.4\linewidth]{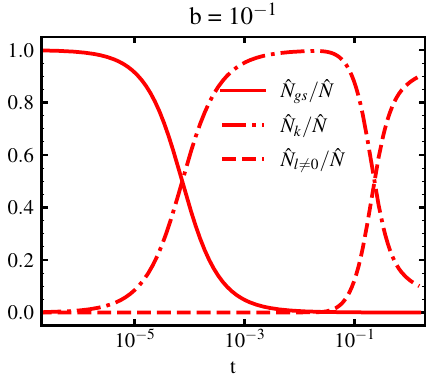} &
\includegraphics[width=0.4\linewidth]{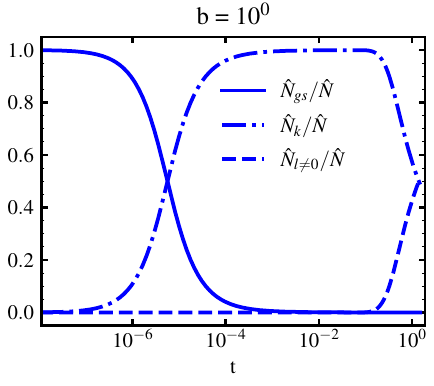} \\
(a) & (b) \\
\multicolumn{2}{c}{
\includegraphics[width=0.4\linewidth]{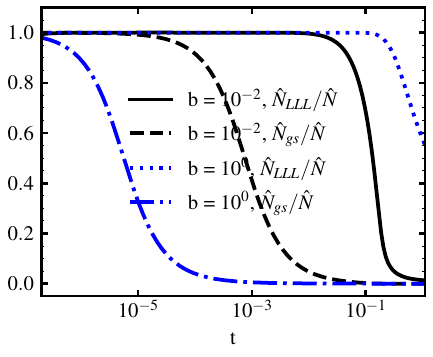}
} \\
\multicolumn{2}{c}{(c)}
\end{tabular}
\caption{Fig. (a) and  (b) show particle population behavior as a function of temperature for fixed values of magnetic field $b=10^{-1}$ and $b=10^{0}$ respectively and density $\hat{N} = 0.012$. Fig (c) shows only the population of $\hat{N}_{LLL}$ and $\hat{N}_{gs}$ as a function of  temperature for different values of the magnetic field.} 
\label{populations}
\end{figure*}
To illustrate the previous analysis on the particle populations and  how the crossover temperatures depend on the magnetic field strength, we have depicted in Fig.~\ref{populations} the relative populations $\hat{N}_{gs}/\hat{N}$, $\hat{N}_{k}/\hat{N}$ and $\hat{N}_{l\ne 0}/\hat{N}$  (with $N_{l\neq0}$ given in Eq.~(\ref{pnumber}c)), \textit{versus} temperature for two values of the magnetic field. 
As the temperature decreases (moving from right to left along the horizontal axis), we observe two key transitions: the first occurs when most particles occupy states with $l = 0$, within $\hat{N}_k$, marking the initial stage of condensation; the second occurs when most particles concentrate near the ground state ($\hat{N}_{gs}$), indicating the final stage of the process. 

For $b = 0.1$ (Fig.~\ref{populations}a), the crossover temperatures are approximately ${t_c}_f \approx 0.07$ and ${t_c}_s \approx 6 \times 10^{-6}$. In contrast, for $b = 1$ (Fig.~\ref{populations}b), the first crossover temperature increases to ${t_c}_f \approx 0.25$, while the second reduces to ${t_c}_s \approx 10^{-6}$. In both cases, in the limit $t \to 0$, we find that $\hat{N}_{gs}/\hat{N} \to 1$. Fig.~\ref{populations}c was depicted to highlight the behavior of $\hat{N}_{gs}$ and $\hat{N}_{LLL}$ as functions of the magnetic field, supporting our previous analysis that the field catalyzes the first condensation step, while it suppresses the second one.

These transition temperatures are also reflected in the specific heat curves shown in Fig.~\ref{fig:cvapp}, and reproduce the two-step BEC described in~Ref.\cite{Delgado2012Twostep} for the non-relativistic MCSBG in a finite volume. Note that the first crossover temperature for $b = 1$ and $b = 0.1$ exactly matches the corresponding one in Fig.~\ref{fig:cvapp}. In contrast, the second crossover temperature obtained from the particle population analysis does not exactly coincide with that in the specific heat plot. This is because, in the particle population analysis, the width $p_0$ of the vicinity around the ground state is arbitrarily chosen.

From this analysis, one concludes that magnetic fields play a dual role in the two-step condensation of the MCSBG. On one hand, it enhances the accumulation of particles in the LLL (because ${t_c}_f$ increases with the magnetic field intensity) effectively reducing the dimensionality of the system and promoting magnetic catalysis. Nevertheless, it simultaneously inhibits the accumulation of particles in the immediate vicinity of the ground state (because $t_{cs}$ decreases with the magnetic field intensity) acting against the second step of the condensation. Besides, the smoothness of the specific heat as a function of the temperature indicates the absence of a second-order phase transition, contrary to the non-magnetized case. However, the changes of the slope of the specific heat per particle curves (manifested as plateaus or shoulders) indicate a two-steps phase transition engined by the accumulation of particles in the LLL and around the ground state as the temperature is decreased.

%%%%%%%%%%%%%%%%%%%%%%%%%%%%%%%%%%%%%%%%%%%%%%%%%%%%%%%%%%%%%%%%%%%%%%%%%%%%%%%%%%%%%%%%%%%%%%%%%%%%%%%%%%%%%%%%%%%%%%%%%%%%%%%

\section{Magnetic properties of the charged boson gas}\label{sec:magn}

This section is devoted to studying the magnetic properties of the MCSBG, considering the contributions from both medium (particles and antiparticles) and vacuum. We start from the thermodynamic potential given by Eq. \eqref{Omega_st_inicial_proper_time}  defining the dimensionless magnetization as~\footnote{To recover the physical dimensions in natural units, this must be multiplied by the factor $em^2$.
}
\begin{align}
    \mathcal{\hat{M}}=- \left( \frac{\partial\hat{\Omega}}{\partial b}   \right)_{t,x}.
\label{vacmagnetism}
\end{align}
 
Since the thermodynamic potential has contributions coming from medium and vacuum, the magnetization can also be decomposed into statistical and vacuum contributions as
 \begin{equation}
 \mathcal{\hat{M}}(t,x,b)=\mathcal{\hat{M}}_{st}(t,x,b)+\mathcal{\hat{M}}_{vac}(b),
 \end{equation}
\noindent
where the vacuum contribution, using Eq.~(\ref{Omegarenor}) in Eq.~(\ref{vacmagnetism}), is given by
\begin{widetext}
\begin{equation}
\mathcal{\hat{M}}_{vac}(b) = -\frac{b }{2 \pi ^2}\zeta^{(1,0)}\left(-1,\frac{b+1}{2 b}\right) 
+\frac{1}{8 \pi ^2}\zeta^{(1,1)}\left(-1,\frac{b+1}{2 b}\right) -\frac{b}{96 \pi ^2} -\frac{1}{32 \pi ^2 b} +\frac{b \log (4)}{96 \pi ^2}+\frac{b \log (b)}{48 \pi ^2}
\label{Mvac}
\end{equation}
\end{widetext}
and is always positive, indicating that the vacuum exhibits a paramagnetic behavior. Additionally, it is a monotonically increasing function of the magnetic field strength.

The statistical magnetization receives contributions from particles ($+$) and antiparticles ($-$) as $\mathcal{\hat{M}}_{st}(t,x,b)= \mathcal{\hat{M}}^+_{st}(t,x,b) + \mathcal{\hat{M}}^-_{st}(t,x,b)$. As before, it can be split into the lowest and excited Landau levels contributions, as
\[
\mathcal{\hat{M}}^\pm_{st}(t,x,b)= \mathcal{\hat{M}}_{LLL}^\pm(t,x,b)+ \mathcal{\hat{M}}_{l\ne 0}^\pm(t,x,b)
\]

We must point out that we first took the derivative with respect to the magnetic field, and only afterwards performed the sum over Landau levels and the low temperature approximation with the Laplace method. We then obtained:
\begin{align}
&\mathcal{\hat{M}}_{LLL}^\pm(t, x,b) 
 =- \dfrac{1}{b}\hat{\Omega}_{LLL}(t, x,b)- \dfrac{1}{2 m_b}\hat{N}_{LLL}(t, x,b), \nonumber \\ 
 &= \frac{t^{1/2}}{4(2 \pi m_b)^{3/2}}\left (-2bm_b Li_{1/2}(z_\pm) + 4 t Li_{{3/2}}(z_\pm)
 \right )  ,   \label{M0}
 \end{align}
 
\noindent and 
\begin{equation}\label{Mlexcit}
\mathcal{\hat{M}}_{l\ne 0} ^\pm (t, x,b)=-m_b^{1/2}\left(\dfrac{t}{2\pi}\right)^{3/2}
\sum\limits_{n=1}^{\infty}\dfrac{z_\pm^n}{n^{3/2}} G(n\gamma),
 \end{equation}
 with 
\begin{equation}
    G(n\gamma)=\dfrac{
1- n\gamma/2- e^{n\gamma}(1 -3/2\gamma n)}{(e^{n\gamma}- 1)^2}.\nonumber
\end{equation}

Additionally, in the regime $m_b \gg t \gg b$, the $l\ne0$ contribution can be approximated as  
\begin{align}
    \mathcal{\hat{M}}_{l\ne 0} ^\pm(&t, x,b)\approx  -m_b^{1/2}\left(\frac{t}{2 \pi}\right)^{3/2} \times \nonumber \\ & \left ( Li_{{3/2}}(z_\pm) -\frac{5}{12} \frac{b}{m_bt} Li_{1/2}(z_\pm)
 \right ).\label{Mlaprox}
\end{align}

Doing $t\rightarrow0$, in the magnetization (\ref{M0}) and (\ref{Mlexcit}) keeping a constant density, one gets
\begin{align}
 \mathcal{\hat{M}}_{st}(t \to 0^+ , x \to m_b^-, b) &=\lim_{t\to 0}\mathcal{\hat{M}}_{LLL}(t, x,b) \nonumber \\ &=  -\frac{\hat{N}}{2 \sqrt{1+b}}.
    \label{t_0_mag}
 \end{align}
 
Since in the limit \( t \to 0 \) all particles are in the condensate (\( \hat{N}_{gs} = \hat{N} \)), the above result indicates that the magnetized scalar condensate exhibits negative statistical magnetization, indicating a \textit{diamagnetic} behavior.
Moreover, even in the limit \( b \to 0 \) the magnetization in dimensional natural units is 
\begin{equation}
\mathcal{M}_{st}(B\to 0^+, T\to 0^+) = -\frac{e}{2m}N, \label{Magrenamente}
\end{equation}
\noindent 
where $e/2m$ is the corresponding Bohr magneton. 
This feature, previously discussed in \cite{ROJAS1996148} is referred to as \textit{ferro-diamagnetism}. 
In contrast, the magnetized vector boson gas exhibits a \textit{ferro-paramagnetic} behavior ~\cite{Quintero2017PRC} because for \( B \to 0 \), the magnetization \( \mathcal{M} \) remains non-zero but positive \cite{QuinteroAngulo2021, Quintero2017PRC}. 

\begin{figure*}[ht!]
\centering
\begin{tabular}{cc}
\includegraphics[width=0.4\linewidth]{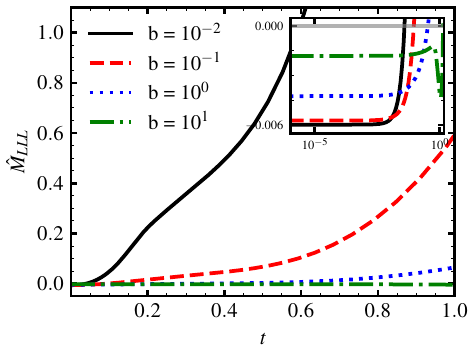} &
\includegraphics[width=0.4\linewidth]{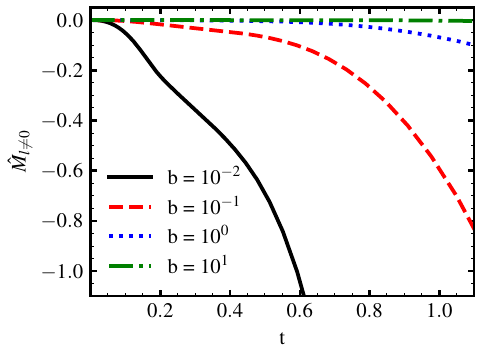} \\
(a) & (b) \\
\end{tabular}
\caption{Contributions to the statistical magnetization as a function of temperature: LLL (a) and excited states $l \ne 0$ (b) at four values of the magnetic field and fixed density.}
\label{fig:MLLL_Ml}
\end{figure*}

Figures (\ref{fig:MLLL_Ml}a) and (\ref{fig:MLLL_Ml}b) show, respectively,  the contributions from the lowest and the highest  Landau levels,  as functions of temperature and constant density. We observe that $\mathcal{\hat{M}}_{LLL}$ transitions from negative to positive values as the temperature increases, with the transition temperature rising with the strength of the magnetic field. For example, at $b = 0.1$ and $b = 10$, the transition temperatures are approximately $0.09$ ($0.09 m_b$) and $1.3$ ($0.39 m_b$), respectively. In contrast $\mathcal{M}_{l\ne 0}$ always remains negative because $G(\gamma n)$ is always positive. Additionally, it shows that the magnetization strength increases with the temperature and decreases with the magnetic field.

\begin{figure*}[ht!]
\centering
\begin{tabular}{cc}
\includegraphics[width=0.4\linewidth]{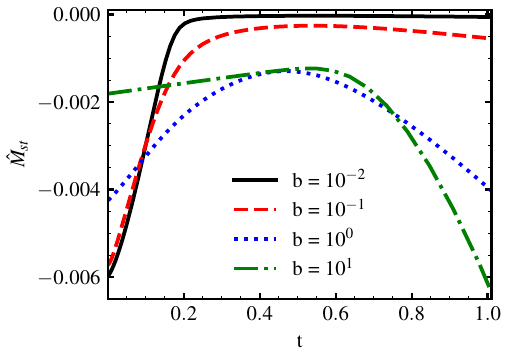}&
\includegraphics[width=0.4\linewidth]{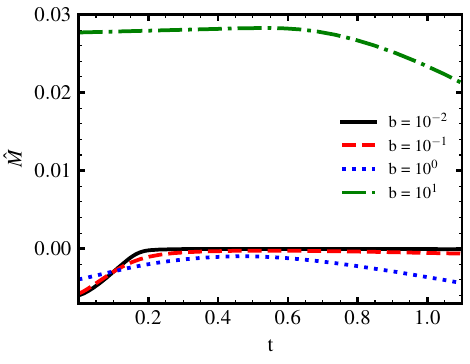}\\
(a) & (b)\\
\end{tabular}
\caption{ Magnetization behavior as function of temperature, at fixed $\hat{N}= 0.012$, (a) only the statistical contribution and (b) the total magnetization: statistical plus vacuum contribution. }
\label{fig:Mbajat}
\end{figure*}

Let us analyze the statistical magnetization as function of temperature displayed in Figure~(\ref{fig:Mbajat}a).
From left to right in the plots, as the temperature increases, the statistical magnetization starts at its zero-temperature value given by Eq. \eqref{t_0_mag}, grows until it reaches a maximum and then diminishes. This maximum flattens for low magnetic fields and diminishes in absolute value; for $b = 0.01$ it nearly disappears and the statistical magnetization is of the order of $10^{-3}$ times the corresponding zero-temperature magnetization. At very low temperatures, the magnetization is higher at stronger magnetic fields and grows faster with the temperature for weak magnetic fields. In contrast, at the highest temperatures, the magnetization is lower and diminishes faster for higher magnetic fields. 

We would like to emphasize that separating the LLL from the states with $l \neq 0$ allows for a better understanding of the system’s magnetization at low temperatures, particularly how the LLL changes the behavior from diamagnetic to paramagnetic. In contrast, the excited states, which always contribute negatively, cause the total magnetization to tend towards zero in weak magnetic fields when the temperature is greater that the zero-magnetic field critical temperature, as pointed out in 
\cite{Schafroth1995, Daicic1996}.
  
Figure~(\ref{fig:Mbajat}b) displays the total magnetization, which includes both the statistical and vacuum contributions. For magnetic fields higher than the critical field (\( b = 10 \)), the total magnetization is dominated by the vacuum contribution at low temperatures, while for magnetic fields \( b \leq 1 \), the statistical contribution is the leading one.

\begin{figure}[h!]
\centering

\includegraphics[width= .9\linewidth]{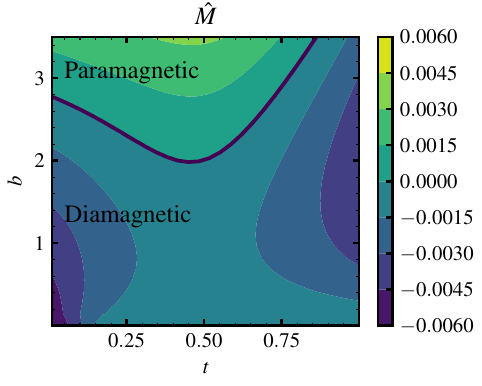}
\caption{Total magnetization: The curve reaches a minimum at \( b \approx 2 \), marking the magnetic field at which paramagnetism appears in the total magnetization. We kept \( \hat{N} = 0.012 \).}
\label{phasetransitionmagnet}
\end{figure}

The interplay between \( \mathcal{\hat{M}}_{st} \) (negative) and \( \mathcal{\hat{M}}_V \) (positive) gives rise to paramagnetic–diamagnetic transitions, characterized by a threshold magnetic field intensity \( b_T \) below which the gas remains diamagnetic at any temperature. In Fig.~\ref{phasetransitionmagnet}, we illustrate the magnetization phase transition in the \( t \)–\( b \) plane at \( \hat{N} = 0.012 \), being $b_T \approx 2$. The region below the curve represents the temperatures and magnetic fields where the magnetization is negative, indicating diamagnetic states of the gas. Conversely, the region above the curve corresponds to temperatures and magnetic fields where the gas exhibits a paramagnetic behavior.

\begin{figure}[ht!]
    \centering
     \includegraphics[width=.9\linewidth]{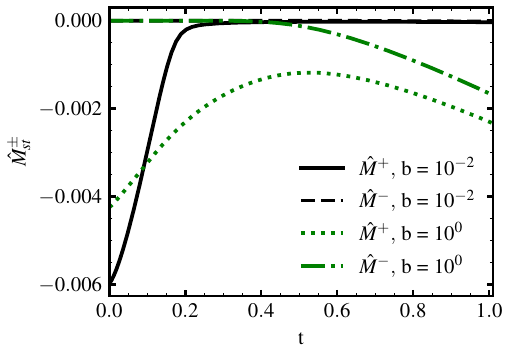}
    \caption{Separated contributions of particles and antiparticles to the magnetization as a function of temperature at two values of the magnetic field and fixed density $\hat{N} = 0.012$. This plot is analog to Fig. (\ref{fig:Mbajat} a) but separating the particles and antiparticles contributions.}
    \label{fig:Mp_Mm}
\end{figure}

Finally, we emphasize the role of antiparticles in the magnetic properties of the system. To illustrate this, we have separately plotted the contributions of particles and antiparticles to the statistical magnetization in Fig. (\ref{fig:Mp_Mm}).
We can see that the particle contribution is dominant, especially at very low temperatures. However, both contributions reach the same order of magnitude at temperatures around $t = 0.8$. Although neglecting the antiparticle contribution would not alter the qualitative behavior of the magnetization, it would lead to an underestimated value and modify the threshold magnetic field. Therefore, antiparticles play a crucial role in the magnetic properties of the system, even in the low-temperature regime of MCSBGs.  

%%%%%%%%%%%%%%%%%%%%%%%%%%%%%%%%%%%%%%%%%%%%%%%%%%%%%%%%%%%%%%%%%%%%%%%%%%%%%%%%%%%%%%%%%%%%%%%%%%%%%%%%%%%%%%%%%%%%%%%%%%%%%%

\section{Equation of state of the MCSBG}\label{sec:EoS}
As the magnetic field breaks the rotational symmetry, the spatial components of the statistical average of the energy-momentum tensor becomes anisotropic, as also occurs in fermion gases \cite{Chaichian:1999gd}. Consequently, the equation of state becomes anisotropic, with the pressure splitting into \textit{parallel} ($\hat{P}_\parallel$) and \textit{perpendicular} ($\hat{P}_\perp$) components, corresponding to the directions parallel and perpendicular to the magnetic field lines. These pressures are
\begin{subequations}
    \begin{align}
        \hat{P}_{\parallel} &=-\hat{\Omega}\\
       \hat{P}_{\perp} &=-\hat{\Omega}-\hat{\mathcal{M}}b,
    \end{align}
\end{subequations}
\noindent where $\hat{\Omega}$ and $\hat{\mathcal{M}}$ are the thermodynamic potential and magnetization (including the vacuum and statistical contributions and including particles and antiparticles). The anisotropy arises from the term $-\mathcal{\hat{M}}b$, which causes the perpendicular pressure to become higher than the parallel one in the diamagnetic regime, and smaller in the paramagnetic regime. 
\begin{figure*}[ht!]
    \centering
    \begin{tabular}{cccc}       \includegraphics[width=0.4\linewidth]{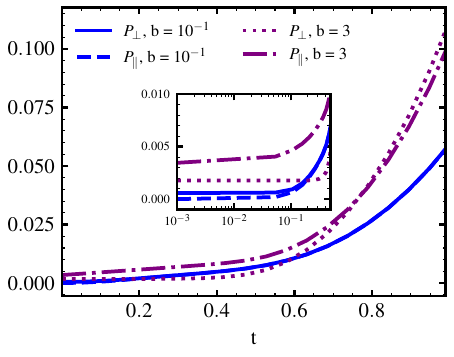} &      \includegraphics[width=0.4\linewidth]{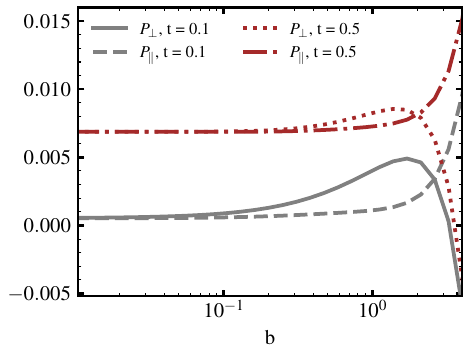} \\
        (a) & (b) \\
        \includegraphics[width=0.4\linewidth]{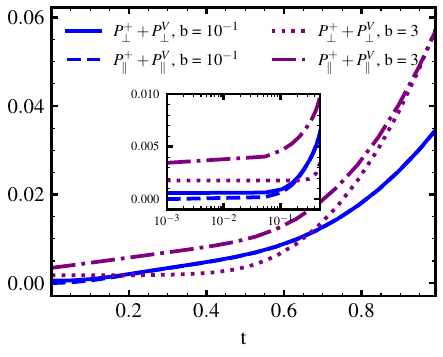} &
\includegraphics[width=0.4\linewidth]
{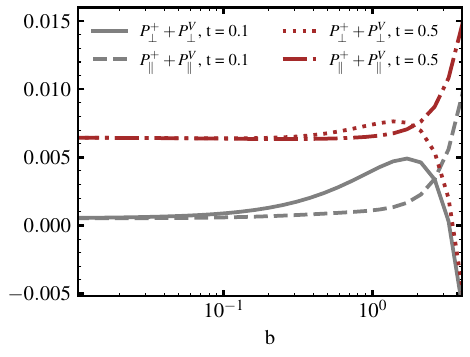}\\
 (c) & (d) \\
    \end{tabular}
    \caption{Parallel and perpendicular pressures (with vacuum and statistical contributions, including particles and antiparticles), (a) as a function of temperature, for two magnetic field strengths; and (b) as a function of magnetic field strength, for two values of the temperature. (c)-(d) Total pressures: statistical ($P_{\parallel}^+$ and $P_{\perp}^+$) plus vacuum pressures ($P_{\parallel}^V$ and $P_{\perp}^V$), (c) as a function of temperature at fixed magnetic field, (d) as a function of the magnetic field at fixed temperature. In these plots the antiparticles contribution to the statistical pressure was  dropped.}   \label{fig:eospressuret}
\end{figure*}

To illustrate the behavior of the pressures at finite density (\(\hat{N} = 0.012\)), Fig.~(\ref{fig:eospressuret}a) shows the pressures as functions of the temperature for two magnetic field strengths. For a weak magnetic field (\(b = 0.1\)), the pressures remain nearly indistinguishable throughout the temperature range. In contrast, for a strong magnetic field (\(b = 3\)), an observable difference is found at all temperatures. For $b = 0.1 < b_T$ the gas is always diamagnetic and $\hat{P}_\parallel < \hat{P}_\perp$. In contrast, for $b = 3 > b_T$, the diamagnetic - paramagnetic regime transition occurs at $t = 0.74$ and, at higher temperatures $\hat{P}_\parallel > \hat{P}_\perp$.

Additionally, in Fig.~(\ref{fig:eospressuret}b), we represent the behavior of the pressures as functions of the magnetic field for two different temperatures. At $t = 0.1$, the pressure splitting becomes evident around $b \gtrsim 10^{-1}$. Moving from left to right (that is, increasing the magnetic field), the perpendicular pressure is the largest at $b < 2.12$, indicating a diamagnetic phase. Initially, it increases, reaching a maximum around $b \approx 1.7$. 
Similarly, at $t = 0.5$, the pressures separate at a higher magnetic field intensity, also in the order of $b \sim 10^{-1} - 1$. The perpendicular pressure is the highest one for $b < 1.7$ and reaches the maximum at $b = 1.38$. In both cases, the gas eventually undergoes a transversal collapse ($\hat{P}_\perp < 0$) as the magnetic field increases. 

In order to evaluate the role of antiparticles, we reproduce in Fig. \eqref{fig:eospressuret}(c-d) the same plots but taking only the contributions from particles and vacuum. Despite this, the qualitative behavior of the curves remains unchanged. Nevertheless, in the pressure \textit{versus} temperature plots, we observe that the pressures decrease in magnitude at $t \gtrsim 0.7$, as this is the temperature range in which the number of antiparticles becomes non-negligible.

%%%%%%%%%%%%%%%%%%%%%%%%%%%%%%%%%%%%%%%%%%%%%%%%%%%%%%%%%%%%%%%%%%%%%%%%%%%%%%%%%%%%%%%%%%%%%%%%%%%%%%%%%%%%%%%%%%%%%%%%%%%%%%

\section{Conclusions} \label{conclusions}

We studied the thermodynamic properties of a relativistic MCSBG in presence of a uniform and constant magnetic field. We focus in the low-temperature regime, for any value of the magnetic field and taking into account both particles and antiparticles within Schwinger's proper-time formalism. This approach allows us to separate the LLL from the excited Landau levels in the low-temperature regime.

The contribution of the LLL to the particle density is divergent when the fugacity $z\to 1$, making the contribution of the LLL dominant in every thermodynamic quantity as the temperature is decreased at constant particle density and magnetic field. The confinement of the bosons in the LLL renders the gas effectively one dimensional. This dimensional reduction from 3D to 1D has been referred to as magnetic catalysis  and is manifested as a plateau in the specific heat curve around $1/2$ in the temperature range in which most particles populate the LLL. 

This dimensional reduction eliminates the second-order transition to the BEC displayed by non-magnetized boson gasses and a diffuse two-step condensation is observed, manifested by the accumulation of a macroscopic number of particles in the LLL (first step) and the vicinity of the ground state of the system (second step). Remarkably, although the magnetic field stimulates the first step of the magnetized Bose-Einstein condensation, it disfavors the second step, because the higher the magnetic field strength, the lower the temperature required to reach the final stage of the condensation. Unlike in the non-magnetized case, a critical temperature cannot be defined for the onset of the Bose-Einstein condensation; instead, crossover temperatures are found related to each step.

The transition from diamagnetism to paramagnetism in the gas occurs when the vacuum contribution becomes the dominant term in the total magnetization. This transition is characterized by a threshold magnetic field, below which the gas remains diamagnetic at any temperature.

For a pion gas at \(10^{12}~\mathrm{g/cm^3}\) the associated threshold magnetic field is \(B_T \approx 2.2 \times 10^{19}~\mathrm{G}\), which is higher than the upper bound of \(\sim 10^{18}~\mathrm{G}\) estimated from the virial theorem for the interior of neutron stars \cite{Duncan1992}. Therefore, if pions are present inside magnetized neutron stars, the gas would behave as diamagnetic.

%The transition from diamagnetism to paramagnetism in the gas occurs when the vacuum contribution becomes the dominant term in the total magnetization. At a density of approximately \(\hat{N} = 0.012\), this transition is characterized by a threshold magnetic field of \(b \approx 2\), below which the gas remains diamagnetic at any temperature.

%For a pion gas, this density corresponds to approximately \(10^{12}~\mathrm{g/cm^3}\), and the associated threshold magnetic field is \(B_T \approx 2.2 \times 10^{19}~\mathrm{G}\). This value is higher than the upper bound of \(\sim 10^{18}~\mathrm{G}\) estimated from the virial theorem for the interior of neutron stars \cite{Duncan1992}. Therefore, if pions are present inside magnetized neutron stars, the gas would behave as diamagnetic.

The EoS is anisotropic, similar to what is observed in fermion gases, and the gas eventually undergoes a magnetic collapse at high magnetic fields. In contrast, at high temperatures, compared to the magnetic field, the anisotropy is negligible. 

Finally, we demonstrated that antiparticles, often neglected in similar studies, contribute significantly even at moderate temperatures. Their inclusion increases the specific heat beyond the classical limit of $3/2$, the total magnetization, and both pressure components. These findings emphasize the importance of relativistic and quantum effects in magnetized bosonic systems, with potential applications not only in neutron star physics but also in laboratory studies of strongly correlated bosonic matter under external fields.

\section{Acknowledgments}
The authors thank Gretel Quintero for the initial discussions that contributed to this work. The work of A.C.A., G.G.P., A.P.M., H.P.R., and A.R.C. was supported by project No. PS211LH012-002 of the Program for Fundamental Research in Physics and Mathematics (IFM), under CITMA, Cuba. G.P.B. was supported by DGAPA-PAPIIT under Grant No. IN117023. A.S. was supported by DGAPA-UNAM under Grant No. PAPIIT-IN108123.

\appendix

%%%%%%%%%%%%%%%%%%%%%%%%%%%%%%%%%%%%%%%%%%%%%%%%%%%%%%%%%%%%%%%%%%%%%%%%%%%%%%%%%%%%%%%%%%%%%%%%%%%%%%%%%%%%%%%%%%%%%%%%%%%%%%%

\section{One-loop effective action for a charged scalar boson}\label{functionaleffective}

In order to show some details to obtain Eq.(\ref{OmegaT}), the thermodynamic potential in the presence of an external magnetic field, let us start by writing the generating functional of connected Feynman diagrams in vacuum, $W[A]$, defined as \cite{LewisH.Ryder282}
\begin{equation}
    e^{i W_{eff}[A] }=\int [\mathcal{D}\Phi^\dagger][\mathcal{D}\Phi]e^{i S[\{\Phi^\dagger(x)\},\{\Phi(x)\},A_\mu]},
\label{effectiveaction}
\end{equation}
where 
\begin{widetext}
\begin{equation}
    S[\{\Phi\},\{\Phi^\dagger\},A_\mu]=\int d^4 x \left((\hat{\Pi}_\mu\Phi(x))^\dagger(\hat{\Pi}^\mu\Phi(x))-m^2\Phi^\dagger(x)\Phi(x)-\frac{1}{4}F^{\mu\nu}F_{\mu\nu}\right)
\end{equation}
\end{widetext}
is the Scalar-QED action, with $\hat{\Pi}^\mu\equiv i(\partial^\mu+ieA^\mu)$, the covariant derivative and $F^{\mu\nu}\equiv\partial^\mu A^\nu-\partial^\nu A^\mu$, the electromagnetic strength tensor.

In order to obtain the one-loop contribution to the effective action, we expand Eq.(\ref{effectiveaction}) up to the linear term in $\hbar$, it is 
\begin{eqnarray}
e^{iW_{eff}[A]}&=&e^{i(W^{0}[A]+W^{1}[A])} 
\end{eqnarray}
with
\begin{equation}
    W^{0}[A]=-\int d^4 x \frac{1}{4}F^{\mu\nu}F_{\mu\nu}
\end{equation}
and
\begin{widetext}
\begin{align}
   e^{i W^{1}[A]}&= e^{i \int d^4x \mathcal{L}^{(1)}} \nonumber \\
   &=\int [\mathcal{D}\Phi^\dagger][\mathcal{D}\Phi]
   e^{i\int d^4 x \left[(\hat{\Pi}_\mu\Phi(x))^\dagger(\hat{\Pi}^\mu\Phi(x))-m^2\Phi^\dagger(x)\Phi(x)\right]}\nonumber \\
   &=\int [\mathcal{D}\Phi^\dagger][\mathcal{D}\Phi]
   e^{i\int d^4 x \ \Phi^\dagger(x)\left[\hat{\Pi}^2-m^2\right]\Phi(x)},
   \label{wone}
   \end{align}
\end{widetext}
where $\mathcal{L}^{(1)}$ is the one-loop effective Lagrangian and in the third line we have integrated by parts.

Since we are dealing with a charged boson in the presence of an external magnetic field, to transform the above equation into the momentum space, we use the Ritus eigenfunction method to represent the boson fields
\begin{equation}
	\Phi(x)=\sum\hspace{-1.3em}\int \frac{d^4p}{(2\pi)^4}\mathbb{E}_{\bar{p}}(x) \phi(\bar{p}),
	\label{ritusep}	
\end{equation} 
with $\mathbb{E}_{\bar{p}}(x)$ the Ritus eigenfunction, which is obtained by solving the Klein-Gordon equation
\begin{equation}
	(\hat{\Pi}^2 - m ^ 2 ) \Phi(x^\mu) = 0.
	\label{Klein-Gordon equation}
\end{equation}
The discrete sum is over the Landau levels and the explicit form of $\mathbb{E}_{\bar{p}}(x)$ and $\bar{p}$ depend on the gauge chosen to describe the external magnetic field.

Thus, using Eq.(\ref{ritusep}) in Eq.(\ref{wone}), we get
\begin{widetext}
\begin{eqnarray}
	e^{i W^{1}[A]}
	&=&\int [\mathcal{D}\phi^\dagger][\mathcal{D}\phi]
	\exp\left\{i\int d^4 x \sum\hspace{-1.5em}\int \frac{d^4p}{(2\pi)^4}\frac{d^4p'}{(2\pi)^4}\mathbb{E}^\dagger_{\bar{p}}(x) \phi^\dagger({\bar{p}}) \left[\hat{\Pi}^2-m^2\right]\mathbb{E}_{\bar{p}'}(x) \phi({\bar{p}}')\right\},\nonumber \\
\end{eqnarray}
\end{widetext}

\noindent
which can be reduced by taking into account the following properties for the $\mathbb{E}_{\bar{p}}(x)$ functions~\cite{ritus_radiative_1972, ritus_method_1978}:
\begin{itemize} 
	\item They diagonalize the canonical momentum $\hat{\Pi}^\mu$
	\begin{equation}
		\hat{\Pi}^\mu \mathbb{E}_p(x) = \bar{p}^\mu \mathbb{E}_p(x)  
	\end{equation}
	\item They are orthogonal
	\begin{equation}
		\int d^{4}x \mathbb{E}_{{\bar{p}}}(x)\mathbb{E}_{{\bar{p}}'}^\dagger(x)
		 =(2\pi)^4\hat{\delta}^{(4)}(\bar{p}-\bar{p}')\label{orthogonality}
	\end{equation}
     \item They satisfy the completeness relation
      \begin{equation}
      	\sum\hspace{-1.3em}\int \frac{d^4p}{(2\pi)^4}\mathbb{E}_{\bar{p}}(x)\mathbb{E}^\dagger_{\bar{p}}(y)=\delta^4(x-y).
      \end{equation}
\end{itemize}

For a magnetic field that defines the $z$-direction, in the asymmetric gauge $A^\mu = (0, 0, Bx, 0)$, the eigenvalue $\bar{p}^\mu$ is given by $\bar{p}^\mu=(p^0,0,\sqrt{(2l+1)eB},p^3)$  with $l = 0, 1, 2, . . .$ denoting the Landau levels; $\hat{\delta}^{(4)}(\bar{p}-\bar{p}')=\delta^{ll'} \delta(p^{0}-p'^{0})\delta(p^{2}-p'^{2}) \delta(p^{3}-p'^{3}) $ and 
\begin{equation}
	\sum\hspace{-1.2em}\int \frac{d^4p}{(2\pi)^4} \equiv \sum_{l=0}^{\infty}\int \frac{dp^0dp^2dp^3}{(2\pi)^4}.
\end{equation}
In this way, we obtain
\begin{align}
	e^{i W^{1}[A]}
	&= \int [\mathcal{D}\phi^\dagger][\mathcal{D}\phi]
	\exp\Bigg\{  i\sum\hspace{-1.5em}\int \frac{d^4p}{(2\pi)^4} \times \nonumber \\ &\phi^\dagger({\bar{p}}) \left[\bar{p}^2-m^2\right]\phi({\bar{p}}) \Bigg\},
\label{functionalW1}
\end{align}
which has a Gaussian form that can be directly computed, yielding \cite{LewisH.Ryder282}
\begin{eqnarray}
	e^{i W^{1}[A]}
	&=& \left(det{\left[\bar{p}^2-m^2\right]}\right)^{-1}
\end{eqnarray}
where $det$ denotes a functional determinant.

Thus, the one loop generating functional of connected Feynman diagrams reads
\begin{align}
	W^{1}[A]
	&= i\ln\left(det{\left[\bar{p}^2-m^2\right]}\right)\nonumber \\
	&= iV \dfrac{eB}{2\pi} \sum_{l=0}^{\infty}\int \frac{dp_0dp_3}{(2\pi)^2} \ln\left(\bar{p}^2-m^2\right),
	\label{OmegaTT}
\end{align}
where $V$ is a 4-dimensional space-time volume. 

Finally, once we take into account the relation between the generating functional $W^1[A]$ and the effective potential, we arrive at~\cite{AshokDas384}
\begin{align}
	\Omega^1[A]=-V^{-1} W^{1}[A].
\end{align}

In the case one deals with a thermodynamic system, which does not evolve in time, in the framework of imaginary time formalism, the effective potential at finite temperature $(T)$ and chemical potential $(\mu)$, that we call along this manuscript  thermodynamic potential, can be obtained by doing the replacements 
	\begin{align}\label{Matsubara}
		p_0-\mu &\rightarrow i(\omega_n+i\mu), \\ 
		\int_{-\infty}^\infty\frac{dp_0}{2\pi} & \rightarrow\frac{1}{\beta}\sum_{\omega_n}, 
	\end{align}
where $n=0,\pm1,\pm2,...$ and $\omega_n\equiv 2n\pi T$ are the Matsubara frequencyies. Then
	\begin{equation}\label{Thermo-Potential}
		\Omega^1[A]=\frac{eB}{\beta}\int_{-\infty}^\infty \frac{dp_3}{4\pi^2}\sum_{l=0}^{\infty}\sum_{\omega_n}\ln \left[(\omega_n+i\mu)^2+E_l^2\right],
	\end{equation}
	where
	\begin{equation}\label{Disp-Rel}
		E_l^2=p_3^2+2|eB|(l+1/2)+m^2,
	\end{equation}
	is the particle spectrum.
	Notice that in this approach, the sum over the $\omega_n$ term, which is obtained in (\ref{Thermo-Potential}), is formally similar to that appearing in the free-particle thermodynamic potential (i.e., at $B=0$ and $\mu\neq 0$) \cite{JosephI.Kapusta385}. Thus, once the sum over Matsubara frequencies is performed, we get

\begin{widetext}
	\begin{equation}\label{Thermo-Potential-2}
		\Omega(T,\mu, B) = \Omega^1[A]=\int_{-\infty}^\infty dp_3\sum_{l=0}^{\infty}\frac{eB}{4\pi^2}\left[E_l+\frac{1}{\beta}\ln \left(1-e^{-\beta(E_l-\mu)}\right)\left(1-e^{-\beta(E_l+\mu)}\right)\right],
	\end{equation}
\end{widetext}
where the factor $\frac{eB}{4\pi^2}$ accounts for the Landau level's degeneracy.

%%%%%%%%%%%%%%%%%%%%%%%%%%%%%%%%%%%%%%%%%%%%%%%%%%%%%%%%%%%%%%%%%%%%%%%%%%%%%%%%%%%%%%%%%%%%%%%%%%%%%%%%%%%%%%%%%%%%%%%%%%%%%%%

\section{Regularization}
\label{potreg}

In order to perform the regularization of the vacuum part of the thermodynamic potential, we shall follow the ideas developed in Ref.~\cite{Schwinger1951}. Let us start by rewriting  Eq.(\ref{vacBpropertime}) as 

\begin{equation}
 \Omega_{\text{V}}(B) = \Omega_{vac}+\Omega_{vac.pol.}+\Omega_V^{R}(B)
	\label{vacBpropertimeAP}.
\end{equation}
where
\begin{align}
	\Omega_{vac}&\equiv -\frac{1}{16\pi^2} \int_{0}^{\infty} \frac{ds}{s^3} e^{-m^2 s} , \\
    \Omega_{vac.pol.}&\equiv -\frac{(eB)^2}{16\pi^2} \int_{0}^{\infty} \frac{ds}{s}  e^{-m^2 s}, \\
\Omega_V^{R}(B)&=-\frac{1}{16\pi^2} \int_{0}^{\infty} \frac{ds}{s^3}  e^{-m^2 s} \times \nonumber \\
    &\Bigg(\dfrac{eBs}{\sinh(eBs)} - 1 + \frac{(eBs)^2}{6}\Bigg).
\end{align}
are the different vacuum contributions to the thermodynamic potential: the first and second terms are identified with vacuum and vacuum polarization, respectively, 
where both are divergent, while the latter is finite~\cite{Schwinger1951}.

To carry out the integration over the proper time on each contribution, we use the identities 
\begin{align}
	\int_0^\infty ds\ s^{a-1} e^{-2hs}&=(2h)^{-\mu}\Gamma(a)  \nonumber \\
    \int_0^\infty ds\ s^{a-1} \frac{e^{-2hs}}{\sinh s}&=     2^{1-a}\Gamma(a)\zeta\left(a,h+\frac{1}{2}\right)
\end{align}
by making a proper selection of the parameters $a$ and $2h$. $\Gamma$ and $\zeta$ are Gamma function and Hurwitz zeta function, respectively. For example, the integral over $s$ for vacuum contribution can be obtained by replacing $\mu\rightarrow 3-\epsilon$, with $\epsilon\rightarrow 0$, and $2h=m^2$, getting
\begin{align}
    \int_{0}^{\infty} \frac{ds}{s^3} e^{-m^2 s}=\frac{1}{\epsilon}-\gamma_E-\ln(m^2)
\end{align}
which gives, as expected, a divergent contribution.

In particular, the regularized vacuum contribution to the thermodynamic potential has the form
\begin{widetext}
\begin{align}
	\Omega_V^{R}(B)&=-\frac{(eB)^{2}}{16\pi^2} \int_{0}^{\infty} ds\ s^{a-1}  e^{-2h s}
	\left(\dfrac{s}{\sinh s} - 1 + \frac{s^2}{6}\right) \nonumber \\
	&=-\frac{(eB)^{2}}{16\pi^2}(2h)^{-a}\Gamma(a)\left(-1+\frac{a(a+1)}{24h^2}+a\ h^a\zeta\left(1+a,\frac{1}{2}+h\right)\right) \nonumber \\ &\stackrel{\mu\rightarrow -2+\epsilon}{=} -\frac{(eB)^{2}}{16\pi^2}
	\left[
	    -\frac{\gamma_E+\ln2h}{6}+h^2(-3+2\gamma_E+\ln(4h^2) +4(-1+\gamma_E+\ln2)\zeta\left(-1,\frac{1}{2}+h\right) -4\zeta^{(1,0)}\left(-1,\frac{1}{2}+h\right)\right]
\end{align}
\end{widetext}
where $2h\equiv B_c/B$, with $B_c=m^2/e$. The technique we have employed is known as $\epsilon$-technique, developed in Ref.\cite{W.Dittrich46}.

%%%%%%%%%%%%%%%%%%%%%%%%%%%%%%%%%%%%%%%%%%%%%%%%%%%%%%%%%%%%%%%%%%%%%%%%%%%%%%%%%%%%%%%%%%%%%%%%%%%%%%%%%%%%%%%%%%%%%%%%%%%%%%%

\section{Bose-Einstein Condensation at $b = 0$}\label{App Bec b = 0}

In this section, we review some key aspects of the usual Bose-Einstein condensation to complete our discussion and facilitate the interpretation of our analysis in the magnetized case. For free non-interacting bosons, the Klein–Gordon equation yields the particle spectrum $E(p) = \sqrt{m^2 + p^2}$ that can be used to compute the thermodynamic potential of the boson gas:
\[
    \Omega_{st}(T, \mu) = - T \int \dfrac{d^3p}{(2\pi)^3} \ln \left[ 1 - e^{\beta(\mu - E(p))} \right],
\]
where we exclude the antiparticle contribution to remain consistent with Section~\ref{sec::BEC}. Making a low-temperature approximation \footnote{In the absence of a magnetic field, a low-temperature approximation is equivalent to use a the non-relativistic spectrum $E(p) = p^2/2m$ as in \cite{Pathria}.}
 ($T \ll m$) the thermodynamic potential reads \cite{Pathria}:
\[
    \hat{\Omega}_{st}(t, x) = - t \left( \dfrac{t}{2\pi} \right)^{3/2} \operatorname{Li}_{5/2}(z),
\]
and the corresponding particle and energy densities are:
\begin{align}
    \hat{N} &= \left( \dfrac{t}{2\pi} \right)^{3/2} \operatorname{Li}_{3/2}(z), \\
    \hat{E} &= \left( \dfrac{t}{2\pi} \right)^{3/2} \left[ Li_{3/2}(z) + \dfrac{3}{2}tLi_{5/2}(z)  \right]
\end{align}
where $z = e^{\beta(x - 1)}$ is the fugacity. Here we start using dimensionless magnitudes for the sake of generality. 

Due to the behavior of the polylogarithmic function $\operatorname{Li}_{3/2}$, the particle density $\hat{N}$ has a temperature-dependent maximum at $z = 1$ ($x = 1$):
\[
    \hat{N}(t,1) = \left( \dfrac{t}{2\pi} \right)^{3/2} \zeta(3/2),
\]
where $\zeta$ is the Riemann zeta function.

If we fix the particle density of the system to a given value $\hat{\rho}$, we can define a critical temperature $t_c$ such that $\hat{\rho} = \hat{N}(t_c, 1)$
and, explicitly:
\[
    t_c = 2\pi \left[ \dfrac{\hat{\rho}}{\zeta(3/2)} \right]^{2/3}.
\]
At any temperature $t \ge t_c$, there exists a value of the chemical potential $x \le 1$ such that $\hat{\rho} = \hat{N}(t, x)$. However, if $t < t_c$, then $\hat{N}(t,1) < \hat{\rho}$ and the gas cannot assimilate all the particles. In this situation, the excess of particles accumulates in the system's ground state ($p = 0$) and the chemical potential becomes equal to the ground-state energy (the boson rest mass), that is $x = 1$ and $z = 1$, $\forall t \le t_c$.

This macroscopic population of the ground state, given by
\[
    \hat{N}_{gs} = \hat{\rho} - \hat{N}(t,1) = \hat{\rho} \left[ 1 - \left( \dfrac{t}{t_c} \right)^{3/2} \right],
\]
is known as the Bose-Einstein condensate. Below the critical temperature, two phases coexist in the system: the BEC with $\hat{N}_{gs}$ particle density, and the gas with $\hat{N}$ particle density. The condensation begins at $t = t_c$ and the population of the ground state increases as the temperature decreases; in particular, at $t = 0$ every particle will be in the BEC, namely $\hat{N}_{gs} = \hat{\rho}$.

The specific heat can be computed by differentiating the energy density with respect to the temperature at fixed $\hat{\rho}$ obtaining \cite{Pathria}

\begin{equation}
    \dfrac{C_V}{\hat{\rho}} \big|_{b = 0} = \begin{cases}\frac{15 Li_{5/2}(z)}{4 Li_{3/2}(z)}-\frac{9 Li_{3/2}(z)}{4 Li_{1/2}(z)}    & t \ge  t_c,\\
        \dfrac{15}{4}\dfrac{\zeta(5/2)}{\zeta(3/2)} \left(\dfrac{t}{t_c}\right)^{3/2},  & t \le  t_c
    \end{cases}
    \label{Cv_b0}
\end{equation}

 This function has a non differentiable maximum at $t =t_c$. Because of this singularity, displayed as a peak in the plots of the specific heat per particle, the phase transition to the Bose-Einstein condensate is classified as a \textit{second-order} of \textit{continuous} phase transition.

\bibliography{bibA4G2}

\begin{thebibliography}{36}%
\makeatletter
\providecommand \@ifxundefined [1]{%
 \@ifx{#1\undefined}
}%
\providecommand \@ifnum [1]{%
 \ifnum #1\expandafter \@firstoftwo
 \else \expandafter \@secondoftwo
 \fi
}%
\providecommand \@ifx [1]{%
 \ifx #1\expandafter \@firstoftwo
 \else \expandafter \@secondoftwo
 \fi
}%
\providecommand \natexlab [1]{#1}%
\providecommand \enquote  [1]{``#1''}%
\providecommand \bibnamefont  [1]{#1}%
\providecommand \bibfnamefont [1]{#1}%
\providecommand \citenamefont [1]{#1}%
\providecommand \href@noop [0]{\@secondoftwo}%
\providecommand \href [0]{\begingroup \@sanitize@url \@href}%
\providecommand \@href[1]{\@@startlink{#1}\@@href}%
\providecommand \@@href[1]{\endgroup#1\@@endlink}%
\providecommand \@sanitize@url [0]{\catcode `\\12\catcode `\$12\catcode `\&12\catcode `\#12\catcode `\^12\catcode `\_12\catcode `\%12\relax}%
\providecommand \@@startlink[1]{}%
\providecommand \@@endlink[0]{}%
\providecommand \url  [0]{\begingroup\@sanitize@url \@url }%
\providecommand \@url [1]{\endgroup\@href {#1}{\urlprefix }}%
\providecommand \urlprefix  [0]{URL }%
\providecommand \Eprint [0]{\href }%
\providecommand \doibase [0]{https://doi.org/}%
\providecommand \selectlanguage [0]{\@gobble}%
\providecommand \bibinfo  [0]{\@secondoftwo}%
\providecommand \bibfield  [0]{\@secondoftwo}%
\providecommand \translation [1]{[#1]}%
\providecommand \BibitemOpen [0]{}%
\providecommand \bibitemStop [0]{}%
\providecommand \bibitemNoStop [0]{.\EOS\space}%
\providecommand \EOS [0]{\spacefactor3000\relax}%
\providecommand \BibitemShut  [1]{\csname bibitem#1\endcsname}%
\let\auto@bib@innerbib\@empty
%</preamble>
\bibitem [{\citenamefont {Chadwick}(1932)}]{Chadwick1932}%
  \BibitemOpen
  \bibfield  {author} {\bibinfo {author} {\bibfnamefont {J.}~\bibnamefont {Chadwick}},\ }\bibfield  {title} {\bibinfo {title} {The existence of a neutron},\ }\href {https://doi.org/10.1098/rspa.1932.0112} {\bibfield  {journal} {\bibinfo  {journal} {Proceedings of the Royal Society of London. Series A, Containing Papers of a Mathematical and Physical Character}\ }\textbf {\bibinfo {volume} {136}},\ \bibinfo {pages} {692} (\bibinfo {year} {1932})}\BibitemShut {NoStop}%
\bibitem [{\citenamefont {Hewish}\ \emph {et~al.}(1968)\citenamefont {Hewish}, \citenamefont {Bell}, \citenamefont {Pilkington}, \citenamefont {Scott},\ and\ \citenamefont {Collins}}]{Hewish1968}%
  \BibitemOpen
  \bibfield  {author} {\bibinfo {author} {\bibfnamefont {A.}~\bibnamefont {Hewish}}, \bibinfo {author} {\bibfnamefont {S.~J.}\ \bibnamefont {Bell}}, \bibinfo {author} {\bibfnamefont {J.~D.~H.}\ \bibnamefont {Pilkington}}, \bibinfo {author} {\bibfnamefont {P.~F.}\ \bibnamefont {Scott}},\ and\ \bibinfo {author} {\bibfnamefont {R.~A.}\ \bibnamefont {Collins}},\ }\bibfield  {title} {\bibinfo {title} {Observation of a rapidly pulsating radio source},\ }\href@noop {} {\bibfield  {journal} {\bibinfo  {journal} {Nature}\ }\textbf {\bibinfo {volume} {217}},\ \bibinfo {pages} {709} (\bibinfo {year} {1968})}\BibitemShut {NoStop}%
\bibitem [{\citenamefont {Kouveliotou}\ \emph {et~al.}(1992)\citenamefont {Kouveliotou}, \citenamefont {Briggs}, \citenamefont {Fishman}, \citenamefont {Dershem}, \citenamefont {Paciesas}, \citenamefont {Pendleton},\ and\ \citenamefont {Hurley}}]{Kouveliotou1992}%
  \BibitemOpen
  \bibfield  {author} {\bibinfo {author} {\bibfnamefont {C.}~\bibnamefont {Kouveliotou}}, \bibinfo {author} {\bibfnamefont {M.~K.}\ \bibnamefont {Briggs}}, \bibinfo {author} {\bibfnamefont {G.~J.}\ \bibnamefont {Fishman}}, \bibinfo {author} {\bibfnamefont {L.~J.}\ \bibnamefont {Dershem}}, \bibinfo {author} {\bibfnamefont {W.~S.}\ \bibnamefont {Paciesas}}, \bibinfo {author} {\bibfnamefont {J.~A.}\ \bibnamefont {Pendleton}},\ and\ \bibinfo {author} {\bibfnamefont {K.}~\bibnamefont {Hurley}},\ }\bibfield  {title} {\bibinfo {title} {Identification of two classes of gamma-ray bursts},\ }\href {https://doi.org/10.1038/355607a0} {\bibfield  {journal} {\bibinfo  {journal} {Nature}\ }\textbf {\bibinfo {volume} {355}},\ \bibinfo {pages} {607} (\bibinfo {year} {1992})}\BibitemShut {NoStop}%
\bibitem [{\citenamefont {Delgado}\ \emph {et~al.}(2012)\citenamefont {Delgado}, \citenamefont {Bargueño},\ and\ \citenamefont {Sols}}]{Delgado2012Twostep}%
  \BibitemOpen
  \bibfield  {author} {\bibinfo {author} {\bibfnamefont {R.~L.}\ \bibnamefont {Delgado}}, \bibinfo {author} {\bibfnamefont {P.}~\bibnamefont {Bargueño}},\ and\ \bibinfo {author} {\bibfnamefont {F.}~\bibnamefont {Sols}},\ }\bibfield  {title} {\bibinfo {title} {Two-step condensation of the charged bose gas},\ }\href {https://doi.org/10.1103/PhysRevE.86.031102} {\bibfield  {journal} {\bibinfo  {journal} {Physical Review E}\ }\textbf {\bibinfo {volume} {86}},\ \bibinfo {pages} {031102} (\bibinfo {year} {2012})},\ \Eprint {https://arxiv.org/abs/arXiv:1207.2093} {arXiv:1207.2093} \BibitemShut {NoStop}%
\bibitem [{\citenamefont {Bigagli}\ \emph {et~al.}(2023)\citenamefont {Bigagli}, \citenamefont {Warner}, \citenamefont {Yuan}, \citenamefont {Zhang}, \citenamefont {Stevenson}, \citenamefont {Karman},\ and\ \citenamefont {Will}}]{Bigagli_2023}%
  \BibitemOpen
  \bibfield  {author} {\bibinfo {author} {\bibfnamefont {N.}~\bibnamefont {Bigagli}}, \bibinfo {author} {\bibfnamefont {C.}~\bibnamefont {Warner}}, \bibinfo {author} {\bibfnamefont {W.}~\bibnamefont {Yuan}}, \bibinfo {author} {\bibfnamefont {S.}~\bibnamefont {Zhang}}, \bibinfo {author} {\bibfnamefont {I.}~\bibnamefont {Stevenson}}, \bibinfo {author} {\bibfnamefont {T.}~\bibnamefont {Karman}},\ and\ \bibinfo {author} {\bibfnamefont {S.}~\bibnamefont {Will}},\ }\bibfield  {title} {\bibinfo {title} {Collisionally stable gas of bosonic dipolar ground-state molecules},\ }\href {https://doi.org/10.1038/s41567-023-02200-6} {\bibfield  {journal} {\bibinfo  {journal} {Nature Physics}\ }\textbf {\bibinfo {volume} {19}},\ \bibinfo {pages} {1579–1584} (\bibinfo {year} {2023})}\BibitemShut {NoStop}%
\bibitem [{\citenamefont {Bigagli}\ \emph {et~al.}(2024)\citenamefont {Bigagli}, \citenamefont {Yuan}, \citenamefont {Zhang}, \citenamefont {Bulatovic}, \citenamefont {Karman}, \citenamefont {Stevenson},\ and\ \citenamefont {Will}}]{Bigagli_2024}%
  \BibitemOpen
  \bibfield  {author} {\bibinfo {author} {\bibfnamefont {N.}~\bibnamefont {Bigagli}}, \bibinfo {author} {\bibfnamefont {W.}~\bibnamefont {Yuan}}, \bibinfo {author} {\bibfnamefont {S.}~\bibnamefont {Zhang}}, \bibinfo {author} {\bibfnamefont {B.}~\bibnamefont {Bulatovic}}, \bibinfo {author} {\bibfnamefont {T.}~\bibnamefont {Karman}}, \bibinfo {author} {\bibfnamefont {I.}~\bibnamefont {Stevenson}},\ and\ \bibinfo {author} {\bibfnamefont {S.}~\bibnamefont {Will}},\ }\bibfield  {title} {\bibinfo {title} {Observation of bose–einstein condensation of dipolar molecules},\ }\href {https://doi.org/10.1038/s41586-024-07492-z} {\bibfield  {journal} {\bibinfo  {journal} {Nature}\ }\textbf {\bibinfo {volume} {631}},\ \bibinfo {pages} {289–293} (\bibinfo {year} {2024})}\BibitemShut {NoStop}%
\bibitem [{\citenamefont {{Rojas}}(1996)}]{ROJAS1996148}%
  \BibitemOpen
  \bibfield  {author} {\bibinfo {author} {\bibfnamefont {H.~P.}\ \bibnamefont {{Rojas}}},\ }\bibfield  {title} {\bibinfo {title} {{Bose-Einstein condensation may occur in a constant magnetic field}},\ }\href {https://doi.org/10.1016/0370-2693(96)00430-3} {\bibfield  {journal} {\bibinfo  {journal} {Physics Letters B}\ }\textbf {\bibinfo {volume} {379}},\ \bibinfo {pages} {148} (\bibinfo {year} {1996})},\ \Eprint {https://arxiv.org/abs/hep-th/9510191} {hep-th/9510191} \BibitemShut {NoStop}%
\bibitem [{\citenamefont {Rojas}(1997)}]{ROJAS1997}%
  \BibitemOpen
  \bibfield  {author} {\bibinfo {author} {\bibfnamefont {H.~P.}\ \bibnamefont {Rojas}},\ }\bibfield  {title} {\bibinfo {title} {On bose-einstein condensation in any dimension},\ }\href {https://doi.org/https://doi.org/10.1016/S0375-9601(97)80679-2} {\bibfield  {journal} {\bibinfo  {journal} {Physics Letters A}\ }\textbf {\bibinfo {volume} {234}},\ \bibinfo {pages} {13 } (\bibinfo {year} {1997})}\BibitemShut {NoStop}%
\bibitem [{\citenamefont {Perez}\ and\ \citenamefont {Villegas}(1999)}]{perez1999boseeinsteincondensationconstantmagnetic}%
  \BibitemOpen
  \bibfield  {author} {\bibinfo {author} {\bibfnamefont {H.}~\bibnamefont {Perez}}\ and\ \bibinfo {author} {\bibfnamefont {L.}~\bibnamefont {Villegas}},\ }\href {https://arxiv.org/abs/cond-mat/9911227} {\bibinfo {title} {Bose-einstein condensation in a constant magnetic field}} (\bibinfo {year} {1999}),\ \Eprint {https://arxiv.org/abs/cond-mat/9911227} {arXiv:cond-mat/9911227 [cond-mat]} \BibitemShut {NoStop}%
\bibitem [{\citenamefont {Daicic}\ and\ \citenamefont {Frankel}(1996)}]{Daicic1996}%
  \BibitemOpen
  \bibfield  {author} {\bibinfo {author} {\bibfnamefont {J.}~\bibnamefont {Daicic}}\ and\ \bibinfo {author} {\bibfnamefont {N.~E.}\ \bibnamefont {Frankel}},\ }\bibfield  {title} {\bibinfo {title} {Superconductivity of the bose gas},\ }\href {https://doi.org/10.1103/PhysRevD.53.5745} {\bibfield  {journal} {\bibinfo  {journal} {Phys. Rev. D}\ }\textbf {\bibinfo {volume} {53}},\ \bibinfo {pages} {5745} (\bibinfo {year} {1996})},\ \bibinfo {note} {published 15 May 1996}\BibitemShut {NoStop}%
\bibitem [{\citenamefont {Khalilov}\ \emph {et~al.}(1997)\citenamefont {Khalilov}, \citenamefont {Ho},\ and\ \citenamefont {Yang}}]{Khalilov1997}%
  \BibitemOpen
  \bibfield  {author} {\bibinfo {author} {\bibfnamefont {V.}~\bibnamefont {Khalilov}}, \bibinfo {author} {\bibfnamefont {C.-L.}\ \bibnamefont {Ho}},\ and\ \bibinfo {author} {\bibfnamefont {C.}~\bibnamefont {Yang}},\ }\bibfield  {title} {\bibinfo {title} {Condensation and magnetization of charged vector boson gas},\ }\href@noop {} {\bibfield  {journal} {\bibinfo  {journal} {Modern Physics Letters A}\ }\textbf {\bibinfo {volume} {12}},\ \bibinfo {pages} {1973} (\bibinfo {year} {1997})}\BibitemShut {NoStop}%
\bibitem [{\citenamefont {Khalilov}\ and\ \citenamefont {Ho}(1999)}]{Khalilov:1999xd}%
  \BibitemOpen
  \bibfield  {author} {\bibinfo {author} {\bibfnamefont {V.~R.}\ \bibnamefont {Khalilov}}\ and\ \bibinfo {author} {\bibfnamefont {C.-L.}\ \bibnamefont {Ho}},\ }\bibfield  {title} {\bibinfo {title} {Pair production of charged vector bosons in supercritical magnetic fields at finite temperatures},\ }\href {https://doi.org/10.1103/PhysRevD.60.033003} {\bibfield  {journal} {\bibinfo  {journal} {Phys. Rev. D}\ }\textbf {\bibinfo {volume} {60}},\ \bibinfo {pages} {033003} (\bibinfo {year} {1999})}\BibitemShut {NoStop}%
\bibitem [{\citenamefont {Standen}\ and\ \citenamefont {Toms}(1999)}]{Standen1999}%
  \BibitemOpen
  \bibfield  {author} {\bibinfo {author} {\bibfnamefont {G.~B.}\ \bibnamefont {Standen}}\ and\ \bibinfo {author} {\bibfnamefont {D.~J.}\ \bibnamefont {Toms}},\ }\bibfield  {title} {\bibinfo {title} {Statistical mechanics of nonrelativistic charged particles in a constant magnetic field},\ }\href {https://doi.org/10.1103/PhysRevE.60.5275} {\bibfield  {journal} {\bibinfo  {journal} {Physical Review E}\ }\textbf {\bibinfo {volume} {60}},\ \bibinfo {pages} {5275} (\bibinfo {year} {1999})}\BibitemShut {NoStop}%
\bibitem [{\citenamefont {Ayala}\ \emph {et~al.}(2012)\citenamefont {Ayala}, \citenamefont {Loewe}, \citenamefont {Rojas},\ and\ \citenamefont {Villavicencio}}]{Ayala:2012dk}%
  \BibitemOpen
  \bibfield  {author} {\bibinfo {author} {\bibfnamefont {A.}~\bibnamefont {Ayala}}, \bibinfo {author} {\bibfnamefont {M.}~\bibnamefont {Loewe}}, \bibinfo {author} {\bibfnamefont {J.~C.}\ \bibnamefont {Rojas}},\ and\ \bibinfo {author} {\bibfnamefont {C.}~\bibnamefont {Villavicencio}},\ }\bibfield  {title} {\bibinfo {title} {{Magnetic catalysis of a charged Bose-Einstein condensate}},\ }\href {https://doi.org/10.1103/PhysRevD.86.076006} {\bibfield  {journal} {\bibinfo  {journal} {Phys. Rev. D}\ }\textbf {\bibinfo {volume} {86}},\ \bibinfo {pages} {076006} (\bibinfo {year} {2012})},\ \Eprint {https://arxiv.org/abs/1208.0390} {arXiv:1208.0390 [hep-ph]} \BibitemShut {NoStop}%
\bibitem [{\citenamefont {Quintero~Angulo}\ \emph {et~al.}(2017)\citenamefont {Quintero~Angulo}, \citenamefont {Pérez~Martínez},\ and\ \citenamefont {Rojas}}]{Quintero2017PRC}%
  \BibitemOpen
  \bibfield  {author} {\bibinfo {author} {\bibfnamefont {G.}~\bibnamefont {Quintero~Angulo}}, \bibinfo {author} {\bibfnamefont {A.}~\bibnamefont {Pérez~Martínez}},\ and\ \bibinfo {author} {\bibfnamefont {H.~P.}\ \bibnamefont {Rojas}},\ }\bibfield  {title} {\bibinfo {title} {{Thermodynamic properties of a neutral vector boson gas in a constant magnetic field}},\ }\href {https://doi.org/10.1103/PhysRevC.96.045810} {\bibfield  {journal} {\bibinfo  {journal} {Phys. Rev.}\ }\textbf {\bibinfo {volume} {C96}},\ \bibinfo {pages} {045810} (\bibinfo {year} {2017})},\ \Eprint {https://arxiv.org/abs/1706.08994} {arXiv:1706.08994 [hep-ph]} \BibitemShut {NoStop}%
%%CITATION = ARXIV:1706.08994;%%
\bibitem [{\citenamefont {Angulo}\ \emph {et~al.}(2021)\citenamefont {Angulo}, \citenamefont {González}, \citenamefont {Martínez},\ and\ \citenamefont {Rojas}}]{QuinteroAngulo2021}%
  \BibitemOpen
  \bibfield  {author} {\bibinfo {author} {\bibfnamefont {G.~Q.}\ \bibnamefont {Angulo}}, \bibinfo {author} {\bibfnamefont {L.~C.~S.}\ \bibnamefont {González}}, \bibinfo {author} {\bibfnamefont {A.~P.}\ \bibnamefont {Martínez}},\ and\ \bibinfo {author} {\bibfnamefont {H.~P.}\ \bibnamefont {Rojas}},\ }\bibfield  {title} {\bibinfo {title} {Magnetized vector boson gas at any temperature},\ }\href {https://doi.org/10.1103/PhysRevC.104.035803} {\bibfield  {journal} {\bibinfo  {journal} {Phys. Rev. C}\ }\textbf {\bibinfo {volume} {104}},\ \bibinfo {pages} {035803} (\bibinfo {year} {2021})}\BibitemShut {NoStop}%
\bibitem [{\citenamefont {Quintero~Angulo}\ \emph {et~al.}(2023)\citenamefont {Quintero~Angulo}, \citenamefont {Su\'arez~Gonz\'alez}, \citenamefont {P\'erez~Mart\'{\i}nez},\ and\ \citenamefont {P\'erez~Rojas}}]{PhysRevC.108.015806}%
  \BibitemOpen
  \bibfield  {author} {\bibinfo {author} {\bibfnamefont {G.}~\bibnamefont {Quintero~Angulo}}, \bibinfo {author} {\bibfnamefont {L.~C.}\ \bibnamefont {Su\'arez~Gonz\'alez}}, \bibinfo {author} {\bibfnamefont {A.}~\bibnamefont {P\'erez~Mart\'{\i}nez}},\ and\ \bibinfo {author} {\bibfnamefont {H.}~\bibnamefont {P\'erez~Rojas}},\ }\bibfield  {title} {\bibinfo {title} {Effects of finite temperature on the magnetized equation of state in neutron stars composed of a bose-einstein condensate},\ }\href {https://doi.org/10.1103/PhysRevC.108.015806} {\bibfield  {journal} {\bibinfo  {journal} {Phys. Rev. C}\ }\textbf {\bibinfo {volume} {108}},\ \bibinfo {pages} {015806} (\bibinfo {year} {2023})}\BibitemShut {NoStop}%
\bibitem [{\citenamefont {Angulo}\ \emph {et~al.}(2017)\citenamefont {Angulo}, \citenamefont {Martinez},\ and\ \citenamefont {P{\'e}rez~Rojas}}]{Angulo:2017hif}%
  \BibitemOpen
  \bibfield  {author} {\bibinfo {author} {\bibfnamefont {G.~Q.}\ \bibnamefont {Angulo}}, \bibinfo {author} {\bibfnamefont {A.~P.}\ \bibnamefont {Martinez}},\ and\ \bibinfo {author} {\bibfnamefont {H.}~\bibnamefont {P{\'e}rez~Rojas}},\ }\bibfield  {title} {\bibinfo {title} {{Condensation of neutral vector bosons with magnetic moment}},\ }\href {https://doi.org/10.1142/S2010194517600473} {\bibfield  {journal} {\bibinfo  {journal} {Int. J. Mod. Phys. Conf. Ser.}\ }\textbf {\bibinfo {volume} {45}},\ \bibinfo {pages} {1760047} (\bibinfo {year} {2017})},\ \Eprint {https://arxiv.org/abs/1701.07916} {arXiv:1701.07916 [hep-ph]} \BibitemShut {NoStop}%
\bibitem [{\citenamefont {Concepci{\'o}n}\ and\ \citenamefont {Angulo}(2024)}]{Concepcion:2024duj}%
  \BibitemOpen
  \bibfield  {author} {\bibinfo {author} {\bibfnamefont {A.~E.~R.}\ \bibnamefont {Concepci{\'o}n}}\ and\ \bibinfo {author} {\bibfnamefont {G.~Q.}\ \bibnamefont {Angulo}},\ }\bibfield  {title} {\bibinfo {title} {{Constraints on Bose{\textendash}Einstein condensate stars as neutron stars models from new observational data}},\ }\href {https://doi.org/10.1002/asna.20240009} {\bibfield  {journal} {\bibinfo  {journal} {Astron. Nachr.}\ }\textbf {\bibinfo {volume} {345}},\ \bibinfo {pages} {e240009} (\bibinfo {year} {2024})},\ \Eprint {https://arxiv.org/abs/2507.08988} {arXiv:2507.08988 [astro-ph.HE]} \BibitemShut {NoStop}%
\bibitem [{\citenamefont {Angulo}\ \emph {et~al.}(2023)\citenamefont {Angulo}, \citenamefont {P\'erez~Mart\'\i{}nez},\ and\ \citenamefont {Rojas}}]{Angulo:2023lil}%
  \BibitemOpen
  \bibfield  {author} {\bibinfo {author} {\bibfnamefont {G.~Q.}\ \bibnamefont {Angulo}}, \bibinfo {author} {\bibfnamefont {A.}~\bibnamefont {P\'erez~Mart\'\i{}nez}},\ and\ \bibinfo {author} {\bibfnamefont {H.~P.}\ \bibnamefont {Rojas}},\ }\bibfield  {title} {\bibinfo {title} {{Bose\textendash{}Einstein condensation of magnetized charged scalar bosons revisited}},\ }\href {https://doi.org/10.1002/asna.20220082} {\bibfield  {journal} {\bibinfo  {journal} {Astron. Nachr.}\ }\textbf {\bibinfo {volume} {344}},\ \bibinfo {pages} {e220082} (\bibinfo {year} {2023})}\BibitemShut {NoStop}%
\bibitem [{\citenamefont {Hern{\'a}ndez}(2025)}]{Hernandez:2025bvn}%
  \BibitemOpen
  \bibfield  {author} {\bibinfo {author} {\bibfnamefont {M.~A.~A.}\ \bibnamefont {Hern{\'a}ndez}},\ }\emph {\bibinfo {title} {{Efecto del campo magn{\'e}tico en estrellas de bosones escalares cargados}}},\ \href@noop {} {\bibinfo {type} {Other thesis}} (\bibinfo {year} {2025}),\ \Eprint {https://arxiv.org/abs/2507.19507} {arXiv:2507.19507 [astro-ph.HE]} \BibitemShut {NoStop}%
\bibitem [{\citenamefont {Ferrer}\ \emph {et~al.}(2015)\citenamefont {Ferrer}, \citenamefont {De~La~Incera}, \citenamefont {Paret}, \citenamefont {Martínez},\ and\ \citenamefont {Sanchez}}]{Ferrer2015AMM}%
  \BibitemOpen
  \bibfield  {author} {\bibinfo {author} {\bibfnamefont {E.~J.}\ \bibnamefont {Ferrer}}, \bibinfo {author} {\bibfnamefont {V.}~\bibnamefont {De~La~Incera}}, \bibinfo {author} {\bibfnamefont {D.~M.}\ \bibnamefont {Paret}}, \bibinfo {author} {\bibfnamefont {A.~P.}\ \bibnamefont {Martínez}},\ and\ \bibinfo {author} {\bibfnamefont {A.}~\bibnamefont {Sanchez}},\ }\bibfield  {title} {\bibinfo {title} {Insignificance of the anomalous magnetic moment of charged fermions for the equation of state of a magnetized and dense medium},\ }\href {https://doi.org/10.1103/PhysRevD.91.085041} {\bibfield  {journal} {\bibinfo  {journal} {Physical Review D}\ }\textbf {\bibinfo {volume} {91}},\ \bibinfo {pages} {085041} (\bibinfo {year} {2015})}\BibitemShut {NoStop}%
\bibitem [{\citenamefont {Gusynin}\ \emph {et~al.}(1996)\citenamefont {Gusynin}, \citenamefont {Miransky},\ and\ \citenamefont {Shovkovy}}]{Gusynin:1995nb}%
  \BibitemOpen
  \bibfield  {author} {\bibinfo {author} {\bibfnamefont {V.~P.}\ \bibnamefont {Gusynin}}, \bibinfo {author} {\bibfnamefont {V.~A.}\ \bibnamefont {Miransky}},\ and\ \bibinfo {author} {\bibfnamefont {I.~A.}\ \bibnamefont {Shovkovy}},\ }\bibfield  {title} {\bibinfo {title} {{Dimensional reduction and catalysis of dynamical symmetry breaking by a magnetic field}},\ }\href {https://doi.org/10.1016/0550-3213(95)00690-A} {\bibfield  {journal} {\bibinfo  {journal} {Nucl. Phys. B}\ }\textbf {\bibinfo {volume} {462}},\ \bibinfo {pages} {249} (\bibinfo {year} {1996})},\ \Eprint {https://arxiv.org/abs/hep-ph/9509320} {arXiv:hep-ph/9509320} \BibitemShut {NoStop}%
\bibitem [{\citenamefont {Chaichian}\ \emph {et~al.}(2000)\citenamefont {Chaichian}, \citenamefont {Masood}, \citenamefont {Montonen}, \citenamefont {Perez~Martinez},\ and\ \citenamefont {Perez~Rojas}}]{Chaichian:1999gd}%
  \BibitemOpen
  \bibfield  {author} {\bibinfo {author} {\bibfnamefont {M.}~\bibnamefont {Chaichian}}, \bibinfo {author} {\bibfnamefont {S.~S.}\ \bibnamefont {Masood}}, \bibinfo {author} {\bibfnamefont {C.}~\bibnamefont {Montonen}}, \bibinfo {author} {\bibfnamefont {A.}~\bibnamefont {Perez~Martinez}},\ and\ \bibinfo {author} {\bibfnamefont {H.}~\bibnamefont {Perez~Rojas}},\ }\bibfield  {title} {\bibinfo {title} {{Quantum magnetic and gravitational collapse}},\ }\href {https://doi.org/10.1103/PhysRevLett.84.5261} {\bibfield  {journal} {\bibinfo  {journal} {Phys. Rev. Lett.}\ }\textbf {\bibinfo {volume} {84}},\ \bibinfo {pages} {5261} (\bibinfo {year} {2000})},\ \Eprint {https://arxiv.org/abs/hep-ph/9911218} {arXiv:hep-ph/9911218} \BibitemShut {NoStop}%
\bibitem [{\citenamefont {Schwinger}(1951)}]{Schwinger1951}%
  \BibitemOpen
  \bibfield  {author} {\bibinfo {author} {\bibfnamefont {J.}~\bibnamefont {Schwinger}},\ }\bibfield  {title} {\bibinfo {title} {On gauge invariance and vacuum polarization},\ }\href {https://doi.org/10.1103/PhysRev.82.664} {\bibfield  {journal} {\bibinfo  {journal} {Physical Review}\ }\textbf {\bibinfo {volume} {82}},\ \bibinfo {pages} {664} (\bibinfo {year} {1951})}\BibitemShut {NoStop}%
\bibitem [{\citenamefont {Dittrich}(2010)}]{W.Dittrich46}%
  \BibitemOpen
  \bibfield  {author} {\bibinfo {author} {\bibfnamefont {W.}~\bibnamefont {Dittrich}},\ }\href@noop {} {\emph {\bibinfo {title} {Probing the Quantum Vacuum: Perturbative Effective Action Approach in Quantum Electrodynamics and Its Application}}},\ Vol.~\bibinfo {volume} {1}\ (\bibinfo  {publisher} {Springer Berlin Heidelberg},\ \bibinfo {year} {2010})\BibitemShut {NoStop}%
\bibitem [{\citenamefont {Arfken}\ and\ \citenamefont {Weber}(2005)}]{Arfken:2005}%
  \BibitemOpen
  \bibfield  {author} {\bibinfo {author} {\bibfnamefont {G.~B.}\ \bibnamefont {Arfken}}\ and\ \bibinfo {author} {\bibfnamefont {H.~J.}\ \bibnamefont {Weber}},\ }\href@noop {} {\emph {\bibinfo {title} {Mathematical Methods for Physicists}}},\ \bibinfo {edition} {6th}\ ed.\ (\bibinfo  {publisher} {Elsevier Academic Press},\ \bibinfo {address} {Amsterdam},\ \bibinfo {year} {2005})\BibitemShut {NoStop}%
\bibitem [{\citenamefont {{Shapiro}}\ and\ \citenamefont {{Teukolsky}}(1983)}]{Shapiro}%
  \BibitemOpen
  \bibfield  {author} {\bibinfo {author} {\bibfnamefont {S.~L.}\ \bibnamefont {{Shapiro}}}\ and\ \bibinfo {author} {\bibfnamefont {S.~A.}\ \bibnamefont {{Teukolsky}}},\ }\href@noop {} {\emph {\bibinfo {title} {Research supported by the National Science Foundation.~New York, Wiley-Interscience, 1983, 663 p.}}}\ (\bibinfo  {publisher} {Wiley},\ \bibinfo {address} {New York, NY},\ \bibinfo {year} {1983})\BibitemShut {NoStop}%
\bibitem [{\citenamefont {Pathria}(1996)}]{Pathria}%
  \BibitemOpen
  \bibfield  {author} {\bibinfo {author} {\bibfnamefont {R.~K.}\ \bibnamefont {Pathria}},\ }\href {http://www.loc.gov/catdir/toc/els032/96001679.html} {{\selectlanguage {English}\emph {\bibinfo {title} {Statistical mechanics / R.K. Pathria}}}},\ \bibinfo {edition} {2nd}\ ed.\ (\bibinfo  {publisher} {Butterworth-Heinemann Oxford ; Boston},\ \bibinfo {year} {1996})\ pp.\ \bibinfo {pages} {xiv, 529 p. :}\BibitemShut {NoStop}%
\bibitem [{\citenamefont {Schafroth}(1955)}]{Schafroth1995}%
  \BibitemOpen
  \bibfield  {author} {\bibinfo {author} {\bibfnamefont {M.~R.}\ \bibnamefont {Schafroth}},\ }\bibfield  {title} {\bibinfo {title} {Superconductivity of a charged ideal bose gas},\ }\href {https://doi.org/10.1103/PhysRev.100.463} {\bibfield  {journal} {\bibinfo  {journal} {Phys. Rev.}\ }\textbf {\bibinfo {volume} {100}},\ \bibinfo {pages} {463} (\bibinfo {year} {1955})}\BibitemShut {NoStop}%
\bibitem [{\citenamefont {Duncan}\ and\ \citenamefont {Thompson}(1992)}]{Duncan1992}%
  \BibitemOpen
  \bibfield  {author} {\bibinfo {author} {\bibfnamefont {R.~C.}\ \bibnamefont {Duncan}}\ and\ \bibinfo {author} {\bibfnamefont {C.}~\bibnamefont {Thompson}},\ }\bibfield  {title} {\bibinfo {title} {Formation of very strongly magnetized neutron stars: Implications for gamma-ray bursts},\ }\href {https://doi.org/10.1086/186413} {\bibfield  {journal} {\bibinfo  {journal} {The Astrophysical Journal}\ }\textbf {\bibinfo {volume} {392}},\ \bibinfo {pages} {L9} (\bibinfo {year} {1992})}\BibitemShut {NoStop}%
\bibitem [{\citenamefont {Ryder}(1996)}]{LewisH.Ryder282}%
  \BibitemOpen
  \bibfield  {author} {\bibinfo {author} {\bibfnamefont {L.~H.}\ \bibnamefont {Ryder}},\ }\href@noop {} {\emph {\bibinfo {title} {Quantum Field Theory}}},\ Vol.~\bibinfo {volume} {1}\ (\bibinfo  {publisher} {Cambridge University Press},\ \bibinfo {year} {1996})\BibitemShut {NoStop}%
\bibitem [{\citenamefont {Ritus}(1972)}]{ritus_radiative_1972}%
  \BibitemOpen
  \bibfield  {author} {\bibinfo {author} {\bibfnamefont {V.~I.}\ \bibnamefont {Ritus}},\ }\bibfield  {title} {\bibinfo {title} {Radiative corrections in quantum electrodynamics with intense field and their analytical properties},\ }\href {https://doi.org/10.1016/0003-4916(72)90191-1} {\bibfield  {journal} {\bibinfo  {journal} {Annals of Physics}\ }\textbf {\bibinfo {volume} {69}},\ \bibinfo {pages} {555} (\bibinfo {year} {1972})}\BibitemShut {NoStop}%
\bibitem [{\citenamefont {Ritus}(1978)}]{ritus_method_1978}%
  \BibitemOpen
  \bibfield  {author} {\bibinfo {author} {\bibfnamefont {V.}~\bibnamefont {Ritus}},\ }\bibfield  {title} {\bibinfo {title} {Method of {Eigenfunctions} and {Mass} {Operator} in {Quantum} {Electrodynamics} of a {Constant} {Field}},\ }\href@noop {} {\bibfield  {journal} {\bibinfo  {journal} {Sov. Phys. JETP}\ }\textbf {\bibinfo {volume} {48}},\ \bibinfo {pages} {788} (\bibinfo {year} {1978})}\BibitemShut {NoStop}%
\bibitem [{\citenamefont {Das}(1997)}]{AshokDas384}%
  \BibitemOpen
  \bibfield  {author} {\bibinfo {author} {\bibfnamefont {A.}~\bibnamefont {Das}},\ }\href@noop {} {\emph {\bibinfo {title} {Finite Temperature Field Theory}}},\ Vol.~\bibinfo {volume} {1}\ (\bibinfo  {publisher} {World Scientific Pub Co Inc},\ \bibinfo {year} {1997})\BibitemShut {NoStop}%
\bibitem [{\citenamefont {Kapusta}\ and\ \citenamefont {Gale}(2006)}]{JosephI.Kapusta385}%
  \BibitemOpen
  \bibfield  {author} {\bibinfo {author} {\bibfnamefont {J.~I.}\ \bibnamefont {Kapusta}}\ and\ \bibinfo {author} {\bibfnamefont {C.}~\bibnamefont {Gale}},\ }\href@noop {} {\emph {\bibinfo {title} {Finite-Temperature Field Theory: Principles and Applications}}},\ Vol.~\bibinfo {volume} {1}\ (\bibinfo  {publisher} {Cambridge University Press},\ \bibinfo {year} {2006})\BibitemShut {NoStop}%
\end{thebibliography}%
\end{document}